# AutoGEEval: A Multimodal and Automated Framework for Geospatial Code Generation on GEE with Large Language Models

Huayi Wu[a,b], Zhangxiao Shen[a], Shuyang Hou[a]*, Jianyuan Liang[a], Haoyue Jiao[c], Yaxian Qing[a], Xiaopu Zhang[a], Xu Li[a], Zhipeng Gui[d], Xuefeng Guan[a], Longgang Xiang[a]

*a. State Key Laboratory of Information Engineering in Surveying, Mapping and Remote Sensing, Wuhan University, Wuhan, China

b. Collaborative Innovation Center of Geospatial Technology, Wuhan University, Wuhan, China

c. School of Resource and Environmental Sciences, Wuhan University, Wuhan, China

d. School of Remote Sensing and Information Engineering, Wuhan University, Wuhan, China

*Corresponding author: Shuyang Hou, email: whuhsy@whu.edu.cn

## Abstract

Geospatial code generation is emerging as a key direction in the integration of artificial intelligence and geoscientific analysis. However, there remains a lack of standardized tools for automatic evaluation in this domain. To address this gap, we propose AutoGEEval, the first multimodal, unit-level automated evaluation framework for geospatial code generation tasks on the Google Earth Engine (GEE) platform powered by large language models (LLMs). Built upon the GEE Python API, AutoGEEval establishes a benchmark suite (AutoGEEval-Bench) comprising 1325 test cases that span 26 GEE data types. The framework integrates both question generation and answer verification components to enable an end-to-end automated evaluation pipeline—from function invocation to execution validation. AutoGEEval supports multidimensional quantitative analysis of model outputs in terms of accuracy, resource consumption, execution efficiency, and error types. We evaluate 18 state-of-the-art LLMs—including general-purpose, reasoning-augmented, code-centric, and geoscience-specialized models—revealing their performance characteristics and potential optimization pathways in GEE code generation. This work provides a unified protocol and foundational resource for the development and assessment of geospatial code generation models, advancing the frontier of automated natural language to domain-specific code translation.

## 1. Introduction

General-purpose code refers to a set of program instructions written in formal languages such as Python, C++, or Java, and is widely applied across diverse tasks including data processing, network communication, and algorithm implementation [1,2]. Through programming, users translate logical intentions into executable tasks on computers [3]. With the rise of Transformer-based large language models (LLMs), models such as GPT-4o, DeepSeek, Claude, and LLaMA have demonstrated remarkable performance in general code generation, owing to their exposure to code patterns in large-scale training corpora and their powerful contextual understanding and generative capabilities [4]. These models allow users to generate code directly from natural language instructions, significantly lowering the entry barrier to programming [5]. Building on this foundation, domain-specific code generation models—such as DeepSeek Coder [6], Qwen2.5-Coder [7], and Code LLaMA [8]—have further improved accuracy and robustness through targeted training. Nevertheless, model-generated code often suffers from issues such as syntax errors, incorrect function calls, or missing dependencies, compromising its executability and logical soundness. This phenomenon, commonly referred to as "code hallucination," remains a challenge [9]. To quantify model performance and guide iterative improvement, researchers have developed benchmark suites such as HumanEval [10], MBPP [11], and LiveCodeBench [12], which enable automated evaluation based on execution success rates and related metrics.

Beyond general-purpose code, the growing reliance on intelligent technologies across disciplines has intensified the demand for task-specific programming tailored to particular data types and analytical workflows. Fields such as biochemistry, finance, and geosciences have successively developed specialized computational platforms—such as Bioconductor [13] and QuantLib [14]—that are typically built upon general-purpose languages (e.g., Python, R, Java) but incorporate deeply customized data structures and processing logic. These platforms have gradually evolved into domain-specific code systems characterized by distinct disciplinary features [15,16]. Compared to general-purpose code, domain-specific code exhibits a high degree of specialization in function naming, parameter definitions, computational logic, and data interfaces, reflecting the reorganization of language semantics and functional customization driven by disciplinary knowledge [17,18]. In geosciences, for instance, the rapid proliferation of high-resolution remote sensing imagery and crowdsourced spatiotemporal data has created escalating demands for tailored geospatial analytical capabilities. In response, cloud-based platforms such as Google Earth Engine (GEE) have emerged as widely adopted, code-driven tools in remote sensing and spatial analysis [19]. GEE offers both JavaScript and Python interfaces, embedding a wide range of geoscientific functions typically prefixed with 'ee.' or 'Export.', which support tasks such as remote sensing preprocessing, index computation, and time-series change detection [20]. Compared to traditional GIS tools dominated by graphical user interfaces (GUIs), GEE's scripting paradigm offers significant advantages in automation and reusability [21]. Users can implement complex workflows via concise scripts and, through its 'copy-paste-run' sharing mechanism, efficiently disseminate geospatial analytical methods, thereby promoting innovation and application across users, regions, and disciplines [22,23].

However, writing code on the GEE platform requires not only basic programming skills but also solid knowledge of geospatial analysis. This includes familiarity with core objects and operators (such as 'ee.Image' and 'ee.FeatureCollection'), remote sensing datasets (e.g., Landsat, MODIS), spatial information concepts (e.g., coordinate systems and geographic projections), and methods for processing and integrating multisource data. As a result, the learning curve for GEE programming is significantly steeper than for general-purpose coding, and users without a geospatial background often encounter substantial barriers in practice [24,25]. With GEE being increasingly applied in domains such as transportation, ecology, and defense, there is a growing demand for more efficient and automated geospatial code generation. In this context, leveraging LLMs to generate GEE code has emerged as a promising approach to lowering the entry barrier and enhancing development efficiency [26]. Existing studies have attempted to construct specialized models, such as GeoCode-GPT [24], by fine-tuning general-purpose LLMs with geospatial code corpora, leading to notable improvements in code quality. However, due to limited training resources, geospatial code accounts for only a small fraction of pretraining data. As a result, models are more prone to "code hallucination" in geospatial code generation tasks than in general domains [27,28]. Typical issues include Function Invocation Errors, Object Type Confusion, Missing Filter Conditions, Loop Structure Errors, Semantic Band Mapping Errors, Type Mismatch Errors, Invalid Type Conversions, and Missing Required Parameters, as illustrated in Figure 1. These issues severely compromise code executability and the reliability of analytical results. Therefore, establishing a systematic evaluation framework for geospatial code generation is essential. It not only helps clarify the performance boundaries of current models in geospatial tasks but also provides theoretical and practical support for developing future high-performance, low-barrier geospatial code generation models [29].

At present, a few studies have begun to explore evaluation mechanisms for geospatial code generation tasks. Representative efforts include GeoCode-Bench [27] and GeoCode-Eval [30] proposed by Wuhan University, as well as the GeoSpatial-Code-LLMs Dataset developed by Wrocław University of Science and Technology [28]. GeoCode-Bench primarily employs multiple-choice, true/false, and open-ended questions. The first two types focus on textual knowledge comprehension without involving actual code generation, while the code-related tasks rely on expert manual scoring, which incurs high evaluation costs, introduces subjectivity, and limits reproducibility. Similarly,

GeoCode-Eval depends on human evaluation and emphasizes complex test cases, lacking systematic testing of basic functions and commonly used logical combinations. This hinders fine-grained analysis of model capabilities. The GeoSpatial-Code-LLMs Dataset attempts to introduce automated evaluation mechanisms but does not yet support multimodal data representations such as imagery, vector, and raster formats. Moreover, its sample size remains limited (approximately 40 instances). In summary, existing evaluation systems exhibit clear limitations in terms of coverage across evaluation dimensions, granularity of assessment, and degree of automation. There is an urgent need to develop an end-to-end, reproducible, and unit-level evaluation benchmark that supports automated assessment and encompasses diverse multimodal geospatial data types.

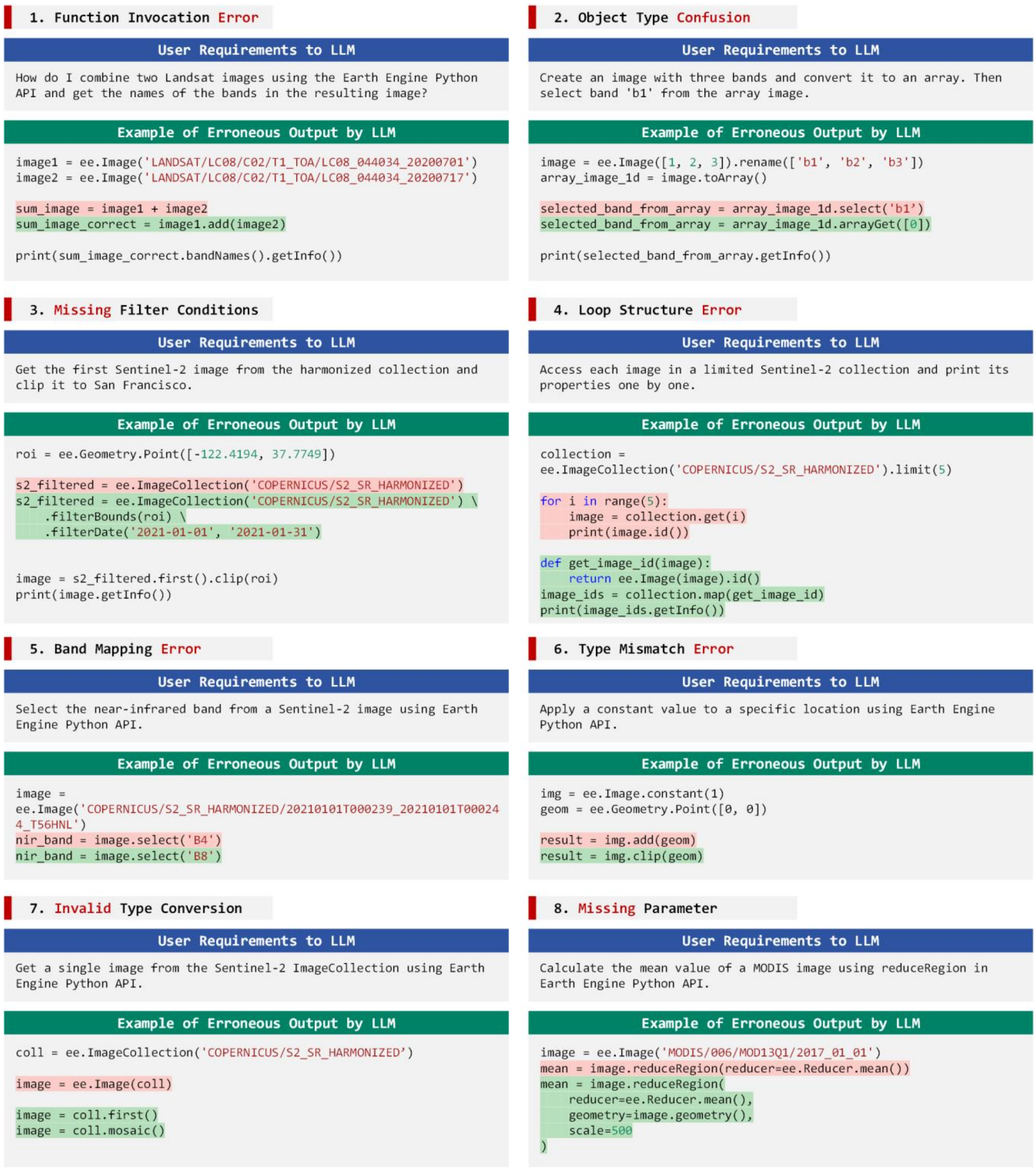

**Figure 1. Common error types in geospatial code generation with LLMs.** The figure categorizes typical errors into three components: the error type, the user's geospatial code prompt to the LLM, and the erroneous code generated by the model. In each example, red highlights the incorrect code, while green indicates the corrected version.

In response to the aforementioned needs and challenges, this study proposes AutoGEEval, an automated evaluation framework for GEE geospatial code generation tasks based on LLMs. The framework supports multimodal data types and unit-level assessment and is implemented using GEE's Python API. On one hand, the Python interface can be executed in local development environments such as Jupyter Notebook and PyCharm, eliminating dependence on the GEE web-based code editor and aligning more closely with real-world development practices. As a result, it has become more widely adopted in applied settings compared to the JavaScript version. On the other hand, Python's local execution environment enables the capture of console outputs and runtime exceptions, thereby facilitating the integration of automated error detection and feedback mechanisms to support a fully end-to-end evaluation workflow. In contrast, the JavaScript interface is constrained by the closed nature of GEE's online platform—its execution process cannot be externally invoked or monitored, making it unsuitable for automation-oriented evaluation tasks.

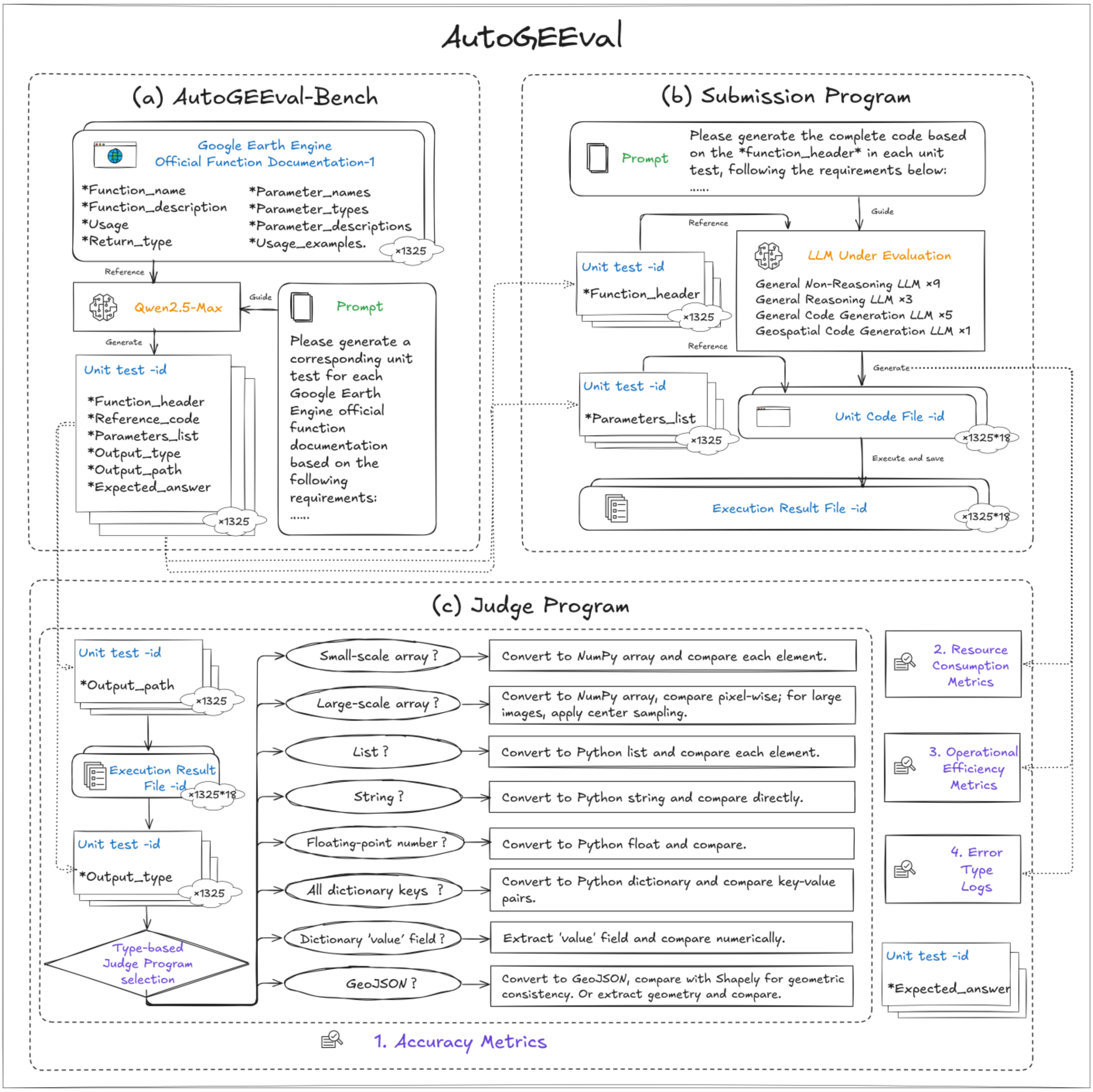

**Figure 2. AutoGEEval framework structure.** The diagram highlights AutoGEEval-Bench, Submission Program, and Judge Program. Blue represents documentation, orange denotes language models, green represents prompts, and purple indicates evaluation metrics.

As illustrated in Figure 2, the AutoGEEval framework consists of three main components: the AutoGEEval-Bench test suite, the Submission Program, and the Judge Program. The AutoGEEval-Bench (see Figure 2a) is constructed based on the official GEE function documentation and contains a total of 1325 unit test cases. All test cases are automatically generated using prompt strategies proposed in this study, guided by the Qwen2.5-Max model, and subsequently verified through rigorous expert review. Each test item comprises six elements: the function declaration (Function_header), a reference code snippet (Reference_code), a list of parameters (Parameters_list), the expected output type (Output_type), the designated output path (Output_path), and the expected answer (Expected_answer). The test suite spans 26 GEE data types, including remote sensing imagery, geometry objects, lists, dictionaries, strings, and numerical values. The Submission Program (see Figure 2b) prompts the target LLM to generate code based on the provided function declaration using carefully designed instructions. It then automatically supplies the required parameters, executes the generated program, and saves the output to the specified path. The Judge Program subsequently reads the output and selects the corresponding evaluation module based on the output type to compute accuracy metrics. In addition, the framework supports automated monitoring and logging of resource consumption, execution efficiency, and error types. In the experimental evaluation, we systematically assessed nine general-purpose non-reasoning models (e.g., GPT-4o), three reasoning-enhanced general models (e.g., DeepSeek-R1), five code generation models (e.g., Qwen2.5-Coder), and one geospatial-specialized model (GeoCode-GPT, including its multi-parameter variants). The results comprehensively reveal performance bottlenecks and potential directions for optimization in current geospatial code generation tasks.

The main contributions of this study are summarized as follows:

- We design, implement, and open-source AutoGEEval, the first automated evaluation framework for geospatial code generation on GEE using LLMs. The framework supports end-to-end automation of test execution, result verification, and error type analysis across multimodal data types at the unit level.
- We construct and release AutoGEEval-Bench, a geospatial code benchmark comprising 1325 unit-level test cases spanning 26 distinct GEE data types.
- We conduct a comprehensive evaluation of 18 representative LLMs across four categories—including GPT-4o, DeepSeek-R1, Qwen2.5-Coder, and GeoCode-GPT—by measuring execution pass rates for geospatial code generation tasks. In addition, we analyze model accuracy, resource consumption, execution efficiency, and error type distributions, providing insights into current limitations and future optimization directions.

The remainder of this paper is organized as follows: Section 2 reviews related work on geospatial code, code generation tasks, and evaluation methods based on LLMs. Section 3 presents the construction methodology of the AutoGEEval-Bench test suite. Section 4 details the design of the AutoGEEval evaluation framework, including the implementation of the Submission and Judge Programs. Section 5 provides a systematic analysis and discussion of the evaluation results. Section 6 concludes the study by summarizing its contributions and significance, identifying current limitations, and outlining future research directions.

## 2. Related work

### 2.1. Geospatial code

Geospatial code refers to a vertical specialization of general-purpose programming languages in the geosciences, specifically denoting code used for processing, analyzing, and visualizing geospatial data. It should be distinguished from terms like "geocoding" or "geospatial encoding," which typically concern the transformation of geographic entities into coordinates or identifiers [24,31-34]. The origins of geospatial code can be traced back to 1963 with the development of the Canada Geographic Information System (CGIS), which introduced batch-processing commands to automate spatial analysis workflows [35]. In 1982, the U.S. Army Corps of Engineers released GRASS GIS [36], which integrated UNIX shell scripting with a modular suite of geographic tools, establishing a code-based paradigm for spatial processing. This was followed by ESRI's ArcInfo and its AML scripting language, which advanced geospatial code toward high-level scripting environments [37]. However, the high maintenance costs and steep learning curves associated with specialized scripting languages limited their broader adoption. In the 1990s, the academic and open-source communities began to embed geospatial processing logic into general-purpose programming languages such as C and Python. The introduction of Perl and Python scripting support in GRASS marked an early integration of GIS tools with mainstream programming languages [38]. Since then, a range of geospatial analysis libraries have been developed for languages including Python, JavaScript, MATLAB, and R, greatly enhancing the accessibility and popularity of spatial analysis [39]. Since 2010, desktop GIS software such as QGIS and ArcGIS have provided Python APIs, facilitating a shift from GUIs to code-driven operations. The launch of GEE marked a new era of cloud-based geospatial computation. Its JavaScript and Python APIs allow for remote access to massive remote sensing datasets and scalable parallel computing frameworks, significantly enhancing geospatial programming capabilities. This evolution has culminated in a modern paradigm where "geospatial code is analysis [40,41]."

### 2.2. Code generation

Early research on code generation focused primarily on general-purpose domains. During the 1980s and 1990s, it relied heavily on handcrafted heuristic rules and template-based systems—such as the pattern-matching mechanism in GCC [42] and tools like Lex/Yacc [43]—which, while efficient, suffered from limited generalizability [44,45]. Since the mid-2010s, with the rise of deep learning, researchers began framing code generation as a sequence-to-sequence problem. Models such as DeepCoder [46] and Seq2SQL [47] achieved impressive performance on specific tasks, but their capabilities remained constrained by limited generative flexibility and dependence on labeled data [48]. Beginning in 2020, the emergence of pretrained and LLMs fundamentally reshaped the paradigm of code generation. Trained on massive corpora of code, these models demonstrated strong instruction-following and contextual reasoning capabilities, enabling end-to-end natural language to code (NL2Code) generation. CodeBERT pioneered dual-modality pretraining across code and natural language [49]. The advent of Codex and Copilot brought NL2Code into practical use [50], while AlphaCode achieved near-human performance in competitive programming [1]. Subsequent models such as DeepSeek-Coder [6] and Qwen2.5-Coder [7] further enriched the technical landscape.

In contrast, geospatial code generation emerged much later and was long constrained by platform-specific, template-based approaches. For example, GEE introduced a parameterized script generator in 2018 [20], and Esri integrated GPT-3-powered code suggestions in 2021 [51]. However, these tools were limited to code completion tasks, lacking generality and adaptability. It was not until October 2024, with the publication of two systematic evaluation papers, that "geospatial code generation" was formally proposed as an independent research task [27,28]. These works extended the general NL2Code [52] paradigm into natural language to geospatial code (NL2GeospatialCode), laying a theoretical foundation for the field. Since then, research in this direction has progressed rapidly. Several optimization strategies have emerged: the CoP strategy employs prompt chaining to guide task decomposition and generation [29]; Geo-FuB [31] and GEE-OPs [25] construct functional semantic and function invocation knowledge bases, respectively, and enhance accuracy through retrieval-augmented generation (RAG); GeoCode-GPT, a Codellama variant fine-tuned on geoscientific code corpora, became the first LLM dedicated to this task [24]. In addition, systems such as MapGPT [53], ShapefileGPT [54], and GIS Copilot [55] integrate tool invocation and knowledge retrieval to enable the automatic generation and execution of geospatial code, signaling the formation of an emerging academic community around this research frontier.

## 2.3. Evaluation of code generation

With the advancement of software engineering and artificial intelligence technologies, evaluation frameworks for general-purpose code generation have evolved from rule-based static analysis toward deeper assessments of semantic understanding and functional correctness [5,56]. Early approaches focused on syntactic rules and basic functional testing. For instance, the GCC compiler released in 1987 provided basic syntax checking but lacked the capability to comprehensively assess code quality [42]. Entering the 2010s, evaluation priorities shifted toward maintainability and structural complexity. Tools such as SonarQube [57] became widely adopted to detect code smells, cyclomatic complexity, and other quality metrics. Industry leaders like Google also established detailed coding standards to enhance code consistency and readability [58]. In parallel, standardized benchmarks for evaluating code generation models began to emerge. Datasets such as HumanEval and MBPP assess models' functional correctness and comprehension capabilities by pairing natural language task descriptions with executable test cases [10,11].

In contrast, evaluating geospatial code generation presents greater challenges, largely due to the complexity of handling multimodal and heterogeneous data (e.g., raster, vector, remote sensing imagery) and the reliance on specific platforms such as GEE or ArcGIS [59,60]. These factors make it difficult to automate evaluation without violating platform usage policies [27,28]. Existing efforts reflect these difficulties. For example, GeoCode-Bench [27] and GeoCode-Eval [24], proposed by Wuhan University, still depend on manual execution and expert judgment, introducing subjectivity into the evaluation process. The University of Wisconsin conducted a preliminary assessment of GPT-4's performance in generating ArcPy code, but neither the evaluation data nor implementation details were publicly released [61]. The GeoSpatial-Code-LLMs Dataset attempts to implement automated evaluation mechanisms, but its scope is limited to basic data types such as GeoDataFrame and Polygon, excluding complex modalities like remote sensing imagery [60]. Moreover, the dataset comprises only about 140 samples, and it lacks diversity in both task complexity and platform environments. As such, its evaluation capability remains limited, leaving ample room for improvement in both breadth and depth.

# 3. AutoGEEval-Bench

AutoGEEval-Bench is constructed based on the official GEE function documentation and consists of 1325 unit-level test cases. All test cases are automatically generated using prompt. The dataset covers 26 GEE data types, including remote sensing imagery, geometry objects, lists, dictionaries, strings, and numerical values. This chapter provides a detailed description of the definition of unit test tasks, the design rationale behind the questions, the construction methodology, and the final composition of the benchmark.

## 3.1. Task definition

Unit-Level Testing ($\mathcal{T}_{\text{unit}}$) is designed to evaluate a model's ability to understand the invocation semantics, parameter structure, and input–output specifications of each API function provided by the platform. The goal is to assess whether the model can generate a syntactically correct and semantically valid function call based on structured function information, such that the code executes successfully and produces the expected result. This task simulates one of the most common workflows for developers—"consulting documentation and writing function calls"—and serves as a capability check at the finest behavioral granularity. Each test case corresponds to a single, independent API function and requires the model to generate executable code that correctly invokes the function with appropriate inputs and yields the expected output.

Let $\mathcal{F}$ denote the set of functions provided in the public documentation of the Earth Engine platform.

$$\mathcal{F} = \{f_1, f_2, \dots, f_N\}, f_i \in GEE_API \quad (1)$$

The task of each model under evaluation is to generate a syntactically correct and executable code snippet $C_i$ within the Earth Engine JavaScript environment.

$$\mathcal{T}_{\text{unit}}: f_i \to C_i \tag{2}$$

Define a code executor, where $y_i$ denotes the result object returned after executing the code snippet $C_i$.

$$\text{Exec}(C_i) = y_i \tag{3}$$

Let $A_i$ denote the expected output (ground-truth answer). The evaluation metric is defined based on the comparison between $y_i$ and $A_i$, where the symbol "=" may represent strict equality, approximate equality for floating-point values, set containment, or other forms of semantic equivalence.

$$\mathcal{L}_{\text{unit}}(y_i, A_i) = \begin{cases} 0, & \text{if Exec}(C_i) = A_i \\ 1, & \text{otherwise} \end{cases} \tag{4}$$

### 3.2. Structural design

All test cases are generated by the flagship LLM Qwen2.5-Max, developed by Alibaba, using predefined prompts and reference data, and subsequently verified by human experts (see Section 3.3 for details). Each complete test case consists of six components: the function header, reference code snippet (Reference_code), parameter list (Parameters_list), output type (Output_type), output path (Output_path), and the expected answer (Expected_answer). Let the set of unit test cases be denoted as:

$$\mathcal{Q} = \{q_1, q_2, \dots, q_n\} \tag{5}$$

Each test case $q_i$ is defined as a six-tuple:

$$q_i = (\mathcal{H}_i, \mathcal{R}_i, \mathcal{P}_i, \mathcal{T}_i, \mathcal{O}_i, \mathcal{A}_i) \tag{6}$$

The meaning of each component is defined as follows:

- $\boldsymbol{\mathcal{H}_i}$ ∈ **FunctionHeader**: Function declaration, including the 'def' statement, function name, parameter list, and a natural language description of the function's purpose. It serves as the semantic prompt to guide the language model in generating the complete function body.
- $\boldsymbol{\mathcal{R}_i}$ ∈ ReferenceCode**:** Reference code snippet, representing the intended logic of the function. It is generated by Qwen2.5-Max based on a predefined prompt and is executed by human experts to obtain the standard answer. During the testing phase, this component is completely hidden from the model, which must independently complete a functionally equivalent implementation based solely on $\mathcal{H}_i$.
- $\boldsymbol{\mathcal{P}_i}$ ∈ ParameterList: Parameter list, specifying the concrete values to be injected into the function during testing, thereby constructing a runnable execution environment.
- $\boldsymbol{\mathcal{T}_i}$ ∈ OutputType**:** Output type, indicating the expected data type returned by the function, used to enforce format constraints on the model's output. Examples include numeric values, Boolean values, dictionaries, or layer objects.
- $\boldsymbol{\mathcal{O}_i}$ ∈ OutputPath**:** Output path, specifying where the execution result of the generated code will be stored. The testing system retrieves the model's output from this path.
- $\boldsymbol{\mathcal{A}_i}$ ∈ ExpectedAnswer**:** Expected answer, the correct output obtained by executing the reference code with the given parameters. It serves as the ground-truth reference for evaluating the accuracy of the model's output.

### 3.3. Construction methodology

The construction of unit test cases is based on the official GEE Reference Documentation, specifically the Client Libraries section, which includes a total of 1,374 functions. Each function page provides the full function name, a description of its functionality, usage examples, return type, parameter names and types, and parameter descriptions.

Some pages include sample code demonstrating function usage, while others do not. An example of the page layout is shown in Figure 3. Prior to constructing the test cases, we manually executed all functions to validate their operability. This process revealed that 43 functions were deprecated or non-functional due to version updates, and were thus excluded. The final set of valid functions incorporated into the unit test suite includes 1325 functions. We extracted relevant information from each function page and organized it into a JSON structure. A corresponding prompt template was then designed (see Figure 4) to guide the LLM in parsing the structured documentation and automatically generating unit-level test items.

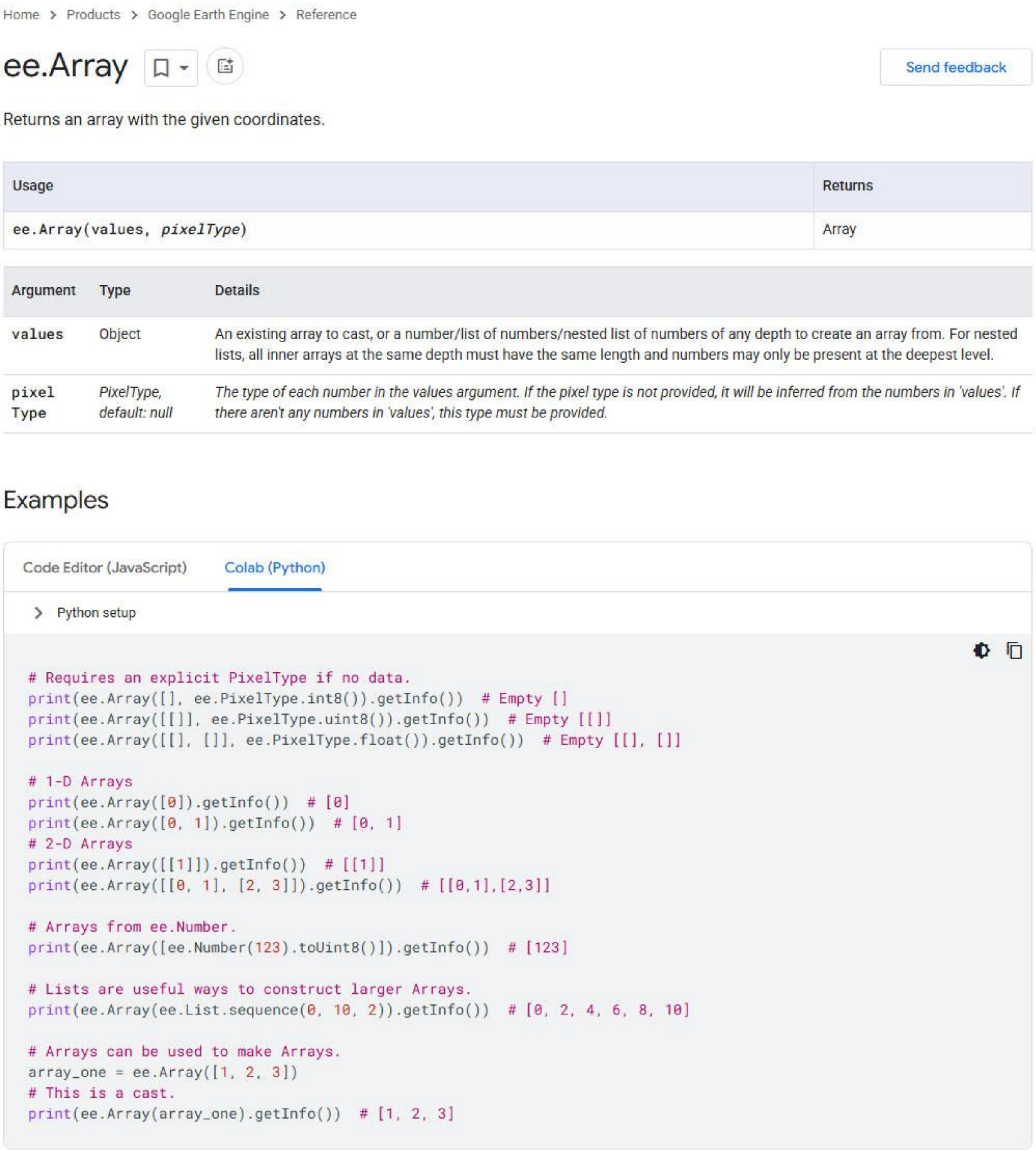

Home > Products > Google Earth Engine > Reference

# ee.Array

Send feedback

Returns an array with the given coordinates.

| Usage | Returns |
|---|---|
| `ee.Array(values, pixelType)` | Array |

| Argument | Type | Details |
|---|---|---|
| values | Object | An existing array to cast, or a number/list of numbers/nested list of numbers of any depth to create an array from. For nested lists, all inner arrays at the same depth must have the same length and numbers may only be present at the deepest level. |
| pixelType | *PixelType, default: null* | *The type of each number in the values argument. If the pixel type is not provided, it will be inferred from the numbers in 'values'. If there aren't any numbers in 'values', this type must be provided.* |

## Examples

Code Editor (JavaScript) | Colab (Python)

> Python setup

```
# Requires an explicit PixelType if no data.
print(ee.Array([], ee.PixelType.int8()).getInfo())  # Empty []
print(ee.Array([[]], ee.PixelType.uint8()).getInfo())  # Empty [[]]
print(ee.Array([[], []], ee.PixelType.float()).getInfo())  # Empty [[], []]

# 1-D Arrays
print(ee.Array([0]).getInfo())  # [0]
print(ee.Array([0, 1]).getInfo())  # [0, 1]
# 2-D Arrays
print(ee.Array([[1]]).getInfo())  # [[1]]
print(ee.Array([[0, 1], [2, 3]]).getInfo())  # [[0,1],[2,3]]

# Arrays from ee.Number.
print(ee.Array([ee.Number(123).toUint8()]).getInfo())  # [123]

# Lists are useful ways to construct larger Arrays.
print(ee.Array(ee.List.sequence(0, 10, 2)).getInfo())  # [0, 2, 4, 6, 8, 10]

# Arrays can be used to make Arrays.
array_one = ee.Array([1, 2, 3])
# This is a cast.
print(ee.Array(array_one).getInfo())  # [1, 2, 3]
```

**Figure 3. Example page from GEE API reference.**

After initial generation, all test cases were manually verified by a panel of five experts with extensive experience in GEE usage and geospatial code development. The verification process ensured that each test task reflects a valid geospatial analysis need, has a clear and accurate problem definition, and is configured with appropriate test inputs.

Any test case exhibiting execution errors or incomplete logic was revised and corrected by the experts based on domain knowledge. The professional backgrounds and selection criteria for these five experts are detailed in Table 1.

**Promt**

```
## Task Description
You need to generate standard test code and configuration file entries for a given Google Earth Engine (GEE) Python API operator.
Each operator will have two parts: the standard code and the test cases in the configuration file.

### Input
1. **Operator Name**: Name of the operator
2. **Explanation**: The explanation of the operator about what it does
3. **Parameter List**: List of parameters with their types and descriptions. For example, `image` (ee.Image): The input image
4. **Return Type**: The return type of the operator

### Output
1. **Standard Code**: Define a function that uses the given operator and returns the result.
The function name should be (Data Type+ operator name + Task). For example, `ee.Image.NormalizedDifference`->`imageNormalizedDifferenceTask`.
2. **Test Cases in Configuration File**: Include multiple test cases, each with parameters, expected answer path, and output type.

### GEE objects in params
1.If the parameter is an GEE object(e.g. ee.Image, ee.Number, etc), use the following format in the configuration file to return the object with python:
param_name: !python |
    def get_ee_object():
        import ee
        ee.Initialize()
        # then get and return the wanted object
2.Notice that some operators may require specific GEE objects as input. e.g. 'ee.Array.CholoskyDecomposition' requires a positive definite ee.Array matrix.

### Output Type
1. The output type can be one of the following:
GEE objects:
"ee.Image", "ee.FeatureCollection", "ee.Number", "ee.List", "ee.Dictionary", "ee.Geometry", "ee.Array", "ee.ImageArray"
Python objects:
"str", "int", "float", "bool", "list", "dict", "NoneType"
2. You can use other types if needed.

### Expected answer
1. The value of the "expected_answer" field in the configuration file MUST be the path to the file containing the expected output.
2. The file name should be (function name + "_testcase" + testcase_number), file type should be .npy for images and arrays,
 .geojson for geometry or feature objects, .txt for other types.

### Note
1. The function should just include ONE operator and return the result. They are used for automatic testing.
2. If the output is a GEE object, do NOT perform getInfo() function. Just return the object.
3. Use the given operator for your answer, do NOT use other methods or operators to solve the task.
4. Any import statements, initialization statements or example usages are NOT needed.
5. Do NOT add any explanation.

### Operator Information
Here is the operator information:
```

**Figure 4. Prompt for unit test construction.**

For test cases that execute successfully and produce the expected results, the output is stored at the specified ‘output_path’ and serves as the ground-truth answer for that item. During the testing phase, the Judge Program retrieves the reference result from this path and compares it against the model-generated output to compute consistency-based accuracy metrics.

**Table 1. Background and selection criteria of experts**

| No. | Age | Qualification | Selection Criteria |
|---|---|---|---|
| 1 | 58 | Professor | A senior PhD supervisor with deep expertise in geospatial services and GIS technology. This expert has authored over 120 publications in internationally recognized journals, overseen several nationally significant research initiatives, and played a leading role in the design and implementation of advanced geospatial analysis platforms. |
| 2 | 35 | Professor | A PhD advisor with a focus on geospatial modeling and intent analysis, specializing in spatiotemporal data mining and GIS applications. This expert serves on the editorial boards of several prestigious international journals and has collaborated with prominent organizations such as NASA, NSF, USGS, GEOSS, and Microsoft. With significant experience in urban transportation systems and the integration of geospatial big data, their expertise spans both theoretical and applied aspects of geospatial technology. |
| 3 | 42 | Algorithm Engineer | A senior algorithm engineer at a prominent internet company, specializing in the development and deployment of LLMs. With a PhD and over 15 years of experience in artificial intelligence and natural language processing, this expert has led numerous large-scale AI projects, particularly focusing on code generation and the application of LLMs across various domains. |
| 4 | 48 | Professor | A PhD advisor and expert in remote sensing with a focus on ecology and agriculture. This specialist has authored numerous interdisciplinary papers published in top international journals. With extensive experience in GEE-based ecological and agricultural remote sensing |

| No. | Age | Qualification | Selection Criteria |
|---|---|---|---|
| | | | technologies, they have received several prestigious national scientific awards for their contributions. |
| 5 | 26 | PhD Candidate | Specializing in geospatial algorithms and code generation, this individual holds both undergraduate and master's degrees in computer science. A first-prize winner in a national GIS analysis competition, they have considerable experience in GEE platform development and algorithm implementation. Their work also includes evaluating code generation from a user-centric perspective, with an emphasis on usability and clarity. |

### 3.4. Construction results

AutoGEEval-Bench includes one corresponding test case for each of the 1325 valid functions officially provided by GEE, resulting in a total of 1325 unit-level testing tasks. These tasks collectively cover 26 different GEE data types, including remote sensing imagery, geometry objects, lists, dictionaries, strings, and numerical values. The distribution and proportion of each data type represented in AutoGEEval-Bench are detailed in Table 2.

**Table 2. Distribution of GEE output types in AutoGEEval-Bench**

| Output_type | Description | Count | Percentage |
|---|---|---|---|
| ee.Array | Multi-dimensional array for numbers and pixels | 118 | 8.91% |
| ee.ArrayImage | Image constructed from multidimensional arrays | 30 | 2.26% |
| ee.Blob | Binary large object storage (e.g., files/models) | 1 | 0.08% |
| ee.BOOL | Boolean logic value (True/False) | 38 | 2.87% |
| ee.Classifier | Machine learning classifier object | 12 | 0.91% |
| ee.Clusterer | Clustering algorithm processor | 6 | 0.45% |
| ee.ConfusionMatrix | Confusion matrix of classification results | 4 | 0.30% |
| ee.Date | Date and time format data | 9 | 0.68% |
| ee.DateRange | Object representing a range of dates | 5 | 0.38% |
| ee.Dictionary | Key-value data structure | 63 | 4.75% |
| ee.Element | Fundamental unit of a geographic feature | 3 | 0.23% |
| ee.ErrorMargin | Statistical object for error margins | 1 | 0.08% |
| ee.Feature | Single feature with properties and shape | 21 | 1.58% |
| ee.FeatureCollection | Collection of geographic features | 41 | 3.09% |
| ee.Filter | Object representing data filtering conditions | 37 | 2.79% |
| ee.Geometry | Geometric shapes (point, line, polygon, etc.) | 146 | 11.02% |
| ee.Image | Single raster image data | 224 | 16.91% |
| ee.ImageCollection | Collection of image data objects | 17 | 1.28% |
| ee.Join | Method for joining datasets | 6 | 0.45% |
| ee.Kernel | Convolution kernel for spatial analysis | 22 | 1.66% |
| ee.List | Ordered list data structure | 68 | 5.13% |
| ee.Number | Numeric data | 194 | 14.64% |
| ee.PixelType | Pixel type definition | 10 | 0.75% |
| ee.Projection | Coordinate system projection information | 15 | 1.13% |
| ee.Reducer | Aggregation and reduction functions | 60 | 4.53% |
| ee.String | String-type data | 174 | 13.13% |
| **Overall** | **Total** | **1325** | **100.00%** |

The 26 GEE data types covered in AutoGEEval-Bench can be broadly categorized into two groups. The first group consists of text-based formats, such as dictionaries, arrays, lists, strings, and floating-point numbers. The second group includes topology-based formats, such as geometries, imagery, and GeoJSON structures. This paper presents representative unit test examples from AutoGEEval-Bench, where Figure 5 illustrates test cases involving text-based GEE data types, while Figure 6 shows examples related to topology-based data types.

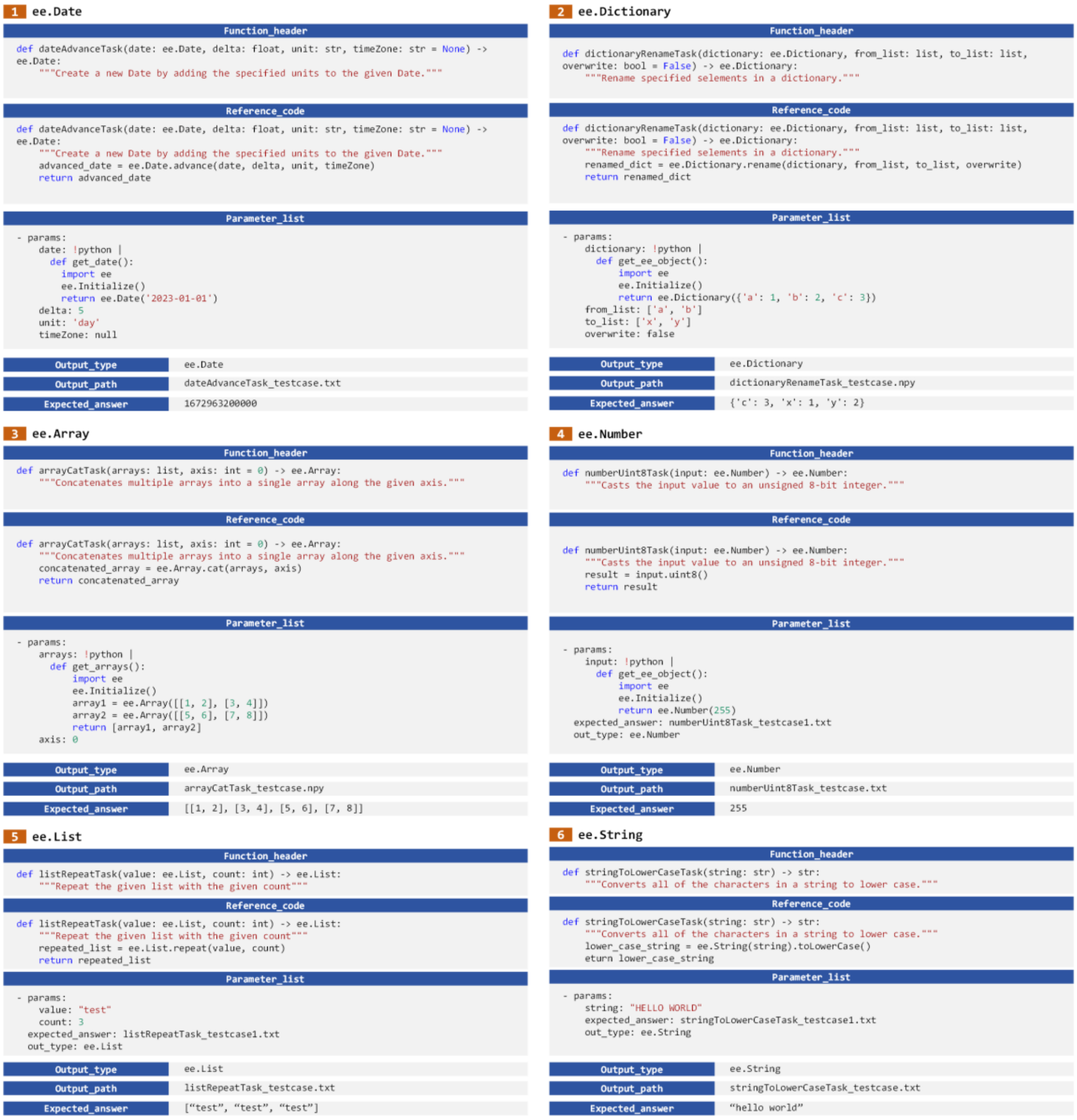

**Figure 5. Unit test example for text-based GEE data types.**

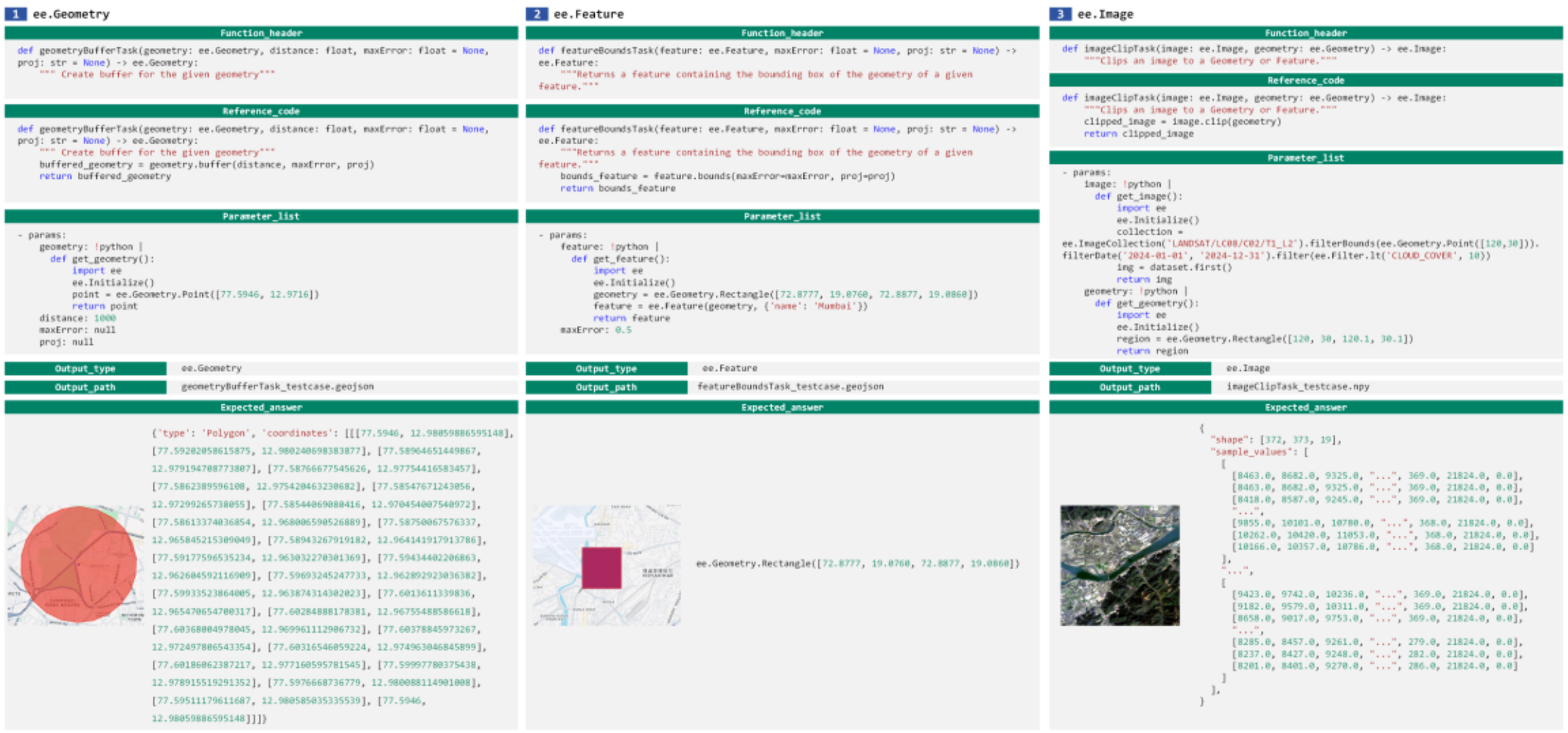

**Figure 6. Unit test example for topology-based GEE data types.**

## 4. Submission and judge programs

During the evaluation phase, the AutoGEEval framework relies on two key components: the Submission Program,

which guides the LLM in responding to the tasks in AutoGEEval-Bench by generating and executing code and saving the results; and the Judge Program, which compares the model's output against the ground-truth answers to determine correctness. This chapter presents the workflow design of both the Submission and Judge Programs.

### 4.1. Submission program

The overall workflow of the Submission Program is illustrated in Figure 2b and consists of three main tasks: answer generation, execution, and result saving. In the answer generation stage, the system utilizes a prompt template to guide the target LLM to respond to each item in AutoGEEval-Bench sequentially. The model generates code based solely on the function header, from which it constructs the corresponding function body. During the execution stage, the execution module reads the parameter list and substitutes the specified values into the formal parameters of the generated code. The code is then executed within the Earth Engine environment. Finally, the execution result is saved to the specified location and file name, as defined by the output path. It is important to note that the prompt is carefully designed to instruct the model to output only the final answer, avoiding any extraneous or irrelevant content. The detailed prompt design is shown in Figure 7.

**Promt**

```
prompt = (
    "Please write the complete GEE Python API function based
on the provided function header. "
    "Only return the function body without any explanations,
comments, or additional text. "
    "The function must use the specified parameters and
produce the expected output. "
    "Ensure that no extra content is included, and do not
modify the function signature or docstring. "
    "Here's the function header and the relevant
information:\n\n"
    f"{test_file_content}"
)
```

**Figure 7. Prompt for submission program**

### 4.2. Judge program

The overall workflow of the Judge Program is illustrated in Figure 2c. Its primary function is to read the execution results from the specified 'Output_path', select the appropriate evaluation logic based on the declared 'Output_type', and compare the model's output against the 'Expected_answer'. The core challenge of the Judge Program lies in accurately assessing correctness across different output data types. As shown in Table 2, AutoGEEval-Bench defines 26 categories of GEE data types. However, many of these types share overlapping numerical representations. For example, although 'ee.Array', 'ee.ConfusionMatrix', and 'ee.ArrayImage' are different in type, they are all expressed as arrays in output. Similarly, 'ee.Dictionary', 'ee.Blob', and 'ee.Reducer' are represented as dictionary-like structures at runtime. Furthermore, 'ee.Geometry', 'ee.Feature', and 'ee.FeatureCollection' all serialize to the GeoJSON format, while both 'ee.String' and 'ee.Boolean' are represented as strings. Given these overlaps, the Judge Program performs unified categorization based on the actual value representation—such as arrays, dictionaries, GeoJSON, or strings—and applies corresponding matching strategies to ensure accurate and fair evaluation across diverse GEE data types. AutoGEEval summarizes the value representations and matching strategies for each GEE data type in Table 3.

**Table 3. Summary of value representations and evaluation strategies for GEE data types**

| GEE Data Type | Value Representation | Testing Logic |
|---|---|---|
| ee.Array | **Small-scale array** | Use getInfo to convert to a NumPy array and compare each element with expected_answer. |
| ee.ConfusionMatrix | | |
| ee.ArrayImage | | |
| ee.Image | **Large-scale array** | Download the image as a NumPy array and perform pixel-wise comparison; for large images, apply center sampling with a |
| ee.ImageCollection | | |

| GEE Data Type | Value Representation | Testing Logic |
| --- | --- | --- |
| | | tolerance of 0.001. Merge all images into one and evaluate as a single image. |
| ee.List | **List** | Convert to a Python list via getInfo and compare each element. |
| ee.String | **String** | Convert to a Python string via getInfo and compare directly. Boolean values are also treated as strings. |
| ee.BOOL | | |
| ee.Number | **Floating-point number** | Convert to a Python float via getInfo and compare with the answer. |
| ee.Dictionary | **All dictionary keys** | Convert to a Python dictionary via getInfo and compare key-value pairs. |
| ee.Blob | | |
| ee.Reducer | | |
| ee.Filter | | |
| ee.Classifier | | |
| ee.Clusterer | | |
| ee.Pixeltype | | |
| ee.Join | | |
| ee.Kernel | | |
| ee.ErrorMargin | | |
| ee.Element | | |
| ee.Projection | | |
| ee.Date | **Dictionary 'value' field** | Use getInfo to obtain a dictionary, extract the 'value' field (timestamp in milliseconds) and compare numerically. |
| ee.DateRange | | |
| ee.Geometry | **GeoJSON** | Convert to GeoJSON using getInfo and compare geometric consistency with Shapely; for Features, extract geometry before comparison. |
| ee.Feature | | |
| ee.FeatureCollection | | |

## 5. Experiments

This chapter outlines the selection criteria for evaluation models, experimental configurations, evaluation metrics, and runtime cost considerations.

### 5.1. Evaluated models

The models evaluated in this study are selected from among the most advanced and widely adopted LLMs as of April 2025. All selected models have either undergone peer review or have been publicly released through open-source or open-access channels. The aim is to align with the growing user preference for end-to-end, easy-to-use models and to provide informative references for both practical application and academic research. It is important to note that optimization strategies such as prompt engineering, RAG, and agent-based orchestration are not included in this evaluation. These strategies do not alter the core model architecture, and their effectiveness is highly dependent on specific design choices, often resulting in unstable performance. Moreover, they are typically tailored for specific downstream tasks and were not originally intended for unit-level testing, making their inclusion in this benchmark neither targeted nor meaningful. Additionally, such strategies often involve complex prompts that consume a large number of tokens, thereby compromising the fairness and efficiency of the evaluation process.

The evaluated models span four categories: (1) general-purpose non-reasoning LLMs, (2) general-purpose reasoning-enhanced LLMs, (3) general-domain code generation models, and (4) task-specific code generation models tailored for geospatial applications. For some models, multiple publicly available parameter configurations are evaluated. Counting different parameter versions as independent models, a total of 18 models are assessed. Detailed specifications of the evaluated models are provided in Table 4.

**Table 4. Information of evaluated LLMs**

| Model Type | Model Name | Developer | Size | Year |
|---|---|---|---|---|
| **General Non-Reasoning** | GPT-4o | OpenAI | N/A | 2024 |
| | GPT-4o-mini | OpenAI | N/A | 2024 |
| | Claude3.7-Sonnet | Anthropic | N/A | 2025 |
| | Gemini-2.0-pro | Google | N/A | 2025 |
| | DeepSeek-V3 | DeepSeek | 671B | 2024 |
| | DeepSeek-V3-0324 | DeepSeek | 685B | 2025 |
| | Qwen-2.5 | Alibaba | 3B, 7B, 32B | 2024 |
| **General Reasoning** | o3-mini | OpenAI | N/A | 2025 |
| | QwQ-32B | Alibaba | 32B | 2025 |
| | DeepSeek-R1 | DeepSeek | 671B | 2025 |
| **General Code Generation** | DeepSeek-Coder-V2 | DeepSeek | 16B | 2024 |
| | Qwen2.5-Coder | Alibaba | 3B, 7B, 32B | 2024 |
| | Code-Llama-7B | Meta | 7B | 2023 |
| **Geospatial Code Generation** | GeoCode-GPT-7B | Wuhan University | 7B | 2024 |

### 5.2. Experimental setup

In terms of hardware configuration and parameter settings, a local computing device equipped with 32GB RAM and an RTX 4090 GPU was used. During model inference, open-source models with parameter sizes not exceeding 16B were deployed locally using the Ollama tool; for larger open-source models and proprietary models, inference was conducted via their official API interfaces to access cloud-hosted versions.

For parameter settings, the generation temperature was set to 0.2 for non-reasoning models to enhance determinism and stability of outputs. For reasoning-enhanced models, following existing research practices, no temperature was specified, preserving the models' native inference capabilities. In addition, the maximum output token length for all models was uniformly set to 4096 to ensure complete responses and prevent truncation due to excessive length.

Time consumption and task descriptions for each phase are provided in Table 5.

**Table 5. Time allocation across experimental stages**

| Stages | Time Spent (hours) |
|---|---|
| AutoGEEval-Bench Construction | 35 |
| Expert Manual Revision | 50 |
| Model Inference and Code Execution | 445 |
| Evaluation of Model Responses | 270 |
| **Total (All Stages)** | **800** |

### 5.3. Evaluation metrics

This study evaluates the performance of LLMs in geospatial code generation tasks across four dimensions: accuracy metrics, resource consumption metrics, operational efficiency metrics, and error type logs.

#### 5.3.1. Accuracy metrics

This study adopts pass@n as the primary accuracy metric. It measures the probability that a correct answer is generated at least once within n independent attempts for the same test case. This is a widely used standard for evaluating both the correctness and stability of model outputs. Given the known hallucination issue in LLMs—where inconsistent or unreliable results may be produced for identical inputs—a single generation may not be representative. Therefore, we evaluate the models under three configurations: n=1,3,5, to enhance the robustness and credibility of the assessment.

$$pass@n = 1 - \frac{C_n}{N} \tag{7}$$

where $N$ is the total number of generated samples, and $C_n$ is the number of incorrect samples among them.
In addition, we introduce the Coefficient of Variation (CV) to assess the stability of the pass@1, pass@3, and pass@5 scores. This metric helps evaluate the variability in model performance across multiple generations, serving as an indirect indicator of the severity of hallucination.

$$CV = \frac{\sigma}{\mu} \tag{8}$$

where $\sigma$ is the standard deviation, and $\mu$ is the mean. A smaller $CV$ indicates higher stability in model performance.
To more comprehensively evaluate model behavior, we further introduce the Stability-Adjusted Accuracy (SA), which integrates both accuracy and stability into a single metric. Specifically, a higher pass@5 score (accuracy) and a lower CV score (stability) result in a higher SA score. The calculation is defined as:

$$SA = \frac{pass@5}{1 + CV} \tag{9}$$

### 5.3.2. Resource consumption metrics

Resource consumption metrics measure the computational resources and cost required for a model to complete the testing tasks. This study considers three key metrics:

- **Token Consumption (Tok.):** Refers to the average number of tokens required to complete each unit test case. For locally deployed models, this metric reflects hardware resource usage; for commercial models, token consumption directly correlates with monetary cost. Most mainstream APIs charge based on the number of tokens processed (typically per 1 million tokens), and pricing varies significantly across models. As of April 2025, GPT-4 Turbo is priced at $10.00/1M tokens, Claude 3 Opus at $15.00/1M tokens, DeepSeek-Coder at $0.60/1M tokens, and Qwen2-72B at $0.80/1M tokens. Therefore, token usage is a critical indicator of both inference cost and model accessibility.
- **Inference Time (In.T):** Refers to the average response time (in seconds) required by the model to generate each test case. This metric reflects latency and response efficiency, both of which directly impact user experience.
- **Code Lines (Co.L):** Measures the number of core executable lines of code generated by the model, excluding comments, natural language explanations, and auxiliary prompts. Compared to token count, code line count provides a more accurate assessment of the model's actual code generation capability, filtering out token inflation caused by unnecessary text in the reasoning process.

### 5.3.3. Operational efficiency metrics

Operational efficiency metrics are used to assess a model's accuracy per unit of resource consumption, thereby reflecting its cost-effectiveness. This study defines inference efficiency, token efficiency, and code line efficiency based on three resource dimensions: time, token usage, and code structure. It is important to note that, to ensure comparability and fairness across models in terms of generation attempts and to reduce the variance caused by random sampling, all resource consumption metrics reported in this study are averaged over five generations. Therefore, pass@5 is uniformly adopted as the reference accuracy metric in all efficiency calculations.

- **Inference Efficiency (In.T-E):** Inference efficiency refers to the average accuracy achieved by a model per unit time, calculated as the ratio of accuracy to average inference time (in seconds). This metric evaluates the model's ability to balance response speed and output quality. The shorter the inference time, the higher the accuracy achieved per unit time, indicating more efficient utilization of computational resources and better interactive performance.

$$Inference\ Efficiency = \frac{pass@5}{Inference\ Time} \tag{10}$$

- **Token Efficiency (Tok.-E):** Token efficiency measures the accuracy achieved per unit of token consumption, calculated as the ratio of accuracy to the average number of tokens used. This metric reflects the economic efficiency of the generation process and effectively supports cross-model comparisons in terms of cost-performance.

$$\text{Token Efficiency} = \frac{pass@5}{\text{Token Consumption}} \quad (11)$$

- **Code Line Efficiency (Co.L-E):** Code line efficiency refers to the accuracy achieved per line of core executable code, emphasizing the structural compactness and effectiveness of the generated logic. Unlike tokens, code lines exclude natural language explanations and prompt-related content, offering a more direct reflection of the model's ability to produce high-quality, executable code for geospatial tasks. This metric is of particular value to developers, especially when evaluating code generation efficiency in practical engineering deployments.

$$\text{Code Efficiency} = \frac{pass@5}{\text{Code Lines}} \quad (12)$$

### 5.3.4. Error type logs

To support qualitative analysis of model performance, AutoGEEval incorporates an automated error detection mechanism based on GEE runtime errors, designed to record the types of errors present in generated code. This mechanism

- **Syntax Error:** These refer to issues in the syntactic structure of the code that prevent successful compilation, such as missing parentheses, misspellings, or missing module imports. Such errors are typically flagged in the GEE console as 'SyntaxError'.
- **Parameter Error:** These occur when the code is syntactically correct but fails to execute due to incorrect or missing parameters. Parameters often involve references to built-in datasets, band names, or other domain-specific knowledge in geosciences. Common error messages include phrases like "xxx has no attribute xx", "xxx not found", or prompts indicating missing required arguments. These errors often arise during parameter concatenation or variable assignment.
- **Invalid Answer:** These refer to cases where the code executes successfully, but the output is inconsistent with the expected answer or the returned data type does not match the predefined specification.
- **Network Error:** These refer to instances where the GEE system returns an Internal Server Error during testing, and the same error persists after three retries under stable network conditions. Such errors are typically caused by logical flaws in the code—such as faulty conditionals or abnormal loop structures—rather than pure syntax issues.

## 6. Results

Building on the evaluation metrics outlined in Section 5.3, this chapter presents a systematic analysis of the evaluation results based on the AutoGEEval framework and the AutoGEEval-Bench test suite. The analysis focuses on four key dimensions: accuracy metrics, resource consumption metrics, operational efficiency metrics, and error type logs.

### 6.1. Accuracy

The evaluation results for accuracy-related metrics across all models are presented in **Table 6**.

**Table 6. Accuracy evaluation results.** Where the values in parentheses under pass@3 represent the improvement over pass@1, and the values in parentheses under pass@5 represent the improvement over pass@3.

| Model | pass@1 (%) | pass@3 (%) | pass@5 (%) | CV | SA |
|---|---|---|---|---|---|
| **General Non-Reasoning** | | | | | |
| GPT-4o | 59.02 | 63.62 (+4.60) | 65.36 (+1.74) | 0.097 | 59.58 |

| Model | pass@1 (%) | pass@3 (%) | pass@5 (%) | CV | SA |
|---|---|---|---|---|---|
| GPT-4o-mini | 55.02 | 60.68 (+4.66) | 61.43 (+0.75) | 0.104 | 55.63 |
| Claude3.7-Sonnet | 63.92 | 66.72 (+2.80) | 67.92 (+1.20) | 0.059 | 64.14 |
| Gemini-2.0-pro | 65.36 | 75.09 (+9.73) | 77.28 (+2.19) | 0.154 | 66.95 |
| DeepSeek-V3 | 71.55 | 75.25 (+3.70) | 76.91 (+1.66) | 0.070 | 71.90 |
| DeepSeek-V3-0324 | 65.28 | 71.92 (+6.64) | 73.51 (+1.59) | 0.112 | 66.11 |
| Qwen-2.5-3B | 33.58 | 39.32 (+5.74) | 41.43 (+2.11) | 0.189 | 34.83 |
| Qwen-2.5-7B | 49.36 | 54.49 (+5.13) | 56.38 (+1.89) | 0.125 | 50.14 |
| Qwen-2.5-32B | 54.42 | 60.00 (+5.58) | 62.04 (+2.04) | 0.123 | 55.25 |
| **General Reasoning** | | | | | |
| o3-mini | 56.98 | 68.91 (+11.93) | 71.02 (+2.11) | 0.198 | 59.30 |
| QwQ-32B | 53.74 | 64.83 (+9.09) | 68.83 (+4.00) | 0.219 | 56.45 |
| DeepSeek-R1 | 60.23 | 72.68 (+12.45) | 76.68 (+4.00) | 0.215 | 63.14 |
| **General Code Generation** | | | | | |
| DeepSeek-Coder-V2 | 31.40 | 38.11 (+6.71) | 40.75 (+2.64) | 0.229 | 33.14 |
| Qwen2.5-Coder-3B | 46.49 | 54.34 (+7.85) | 57.36 (+3.02) | 0.190 | 48.22 |
| Qwen2.5-Coder-7B | 51.25 | 57.66 (+6.41) | 60.91 (+3.25) | 0.159 | 52.57 |
| Qwen2.5-Coder-32B | 61.28 | 64.08 (+2.80) | 65.21 (+1.13) | 0.060 | 61.50 |
| Code-Llama-7B | 56.98 | 64.00 (+7.02) | 66.42 (+2.42) | 0.142 | 58.15 |
| **Geospatial Code Generation** | | | | | |
| GeoCode-GPT-7B | 58.58 | 65.34 (+6.76) | 68.53 (+3.19) | 0.145 | 59.84 |

The stacked bar chart of execution accuracy across models is shown in Figure 8. As observed, increasing the number of generation attempts generally improves accuracy, indicating that multiple generations can partially mitigate hallucination in model-generated code. However, a visual analysis reveals that although both pass@3 and pass@5 increase the number of generations by two rounds compared to the previous level, the green segment (representing the improvement from pass@3 to pass@5) is noticeably shorter than the orange segment (representing the improvement from pass@1 to pass@3). This suggests a significant diminishing return in accuracy gains with additional generations. Quantitative analysis results are presented in Figure 9. The average improvement in pass@3 is 12.88%, ranging from 4.38% to 21.37%. In contrast, the average improvement in pass@5 is only 3.81%, with a range of 1.24% to 6.93%. This pattern highlights a clear diminishing marginal effect in improving accuracy through additional generations. It suggests that while early rounds of generation can substantially correct errors and enhance accuracy, the potential for improvement gradually tapers off in later rounds, reducing the value of further sampling. Therefore, future research should focus on enhancing performance during the initial generation rounds, rather than relying on incremental gains from additional sampling, in order to improve generation efficiency and accuracy more effectively.

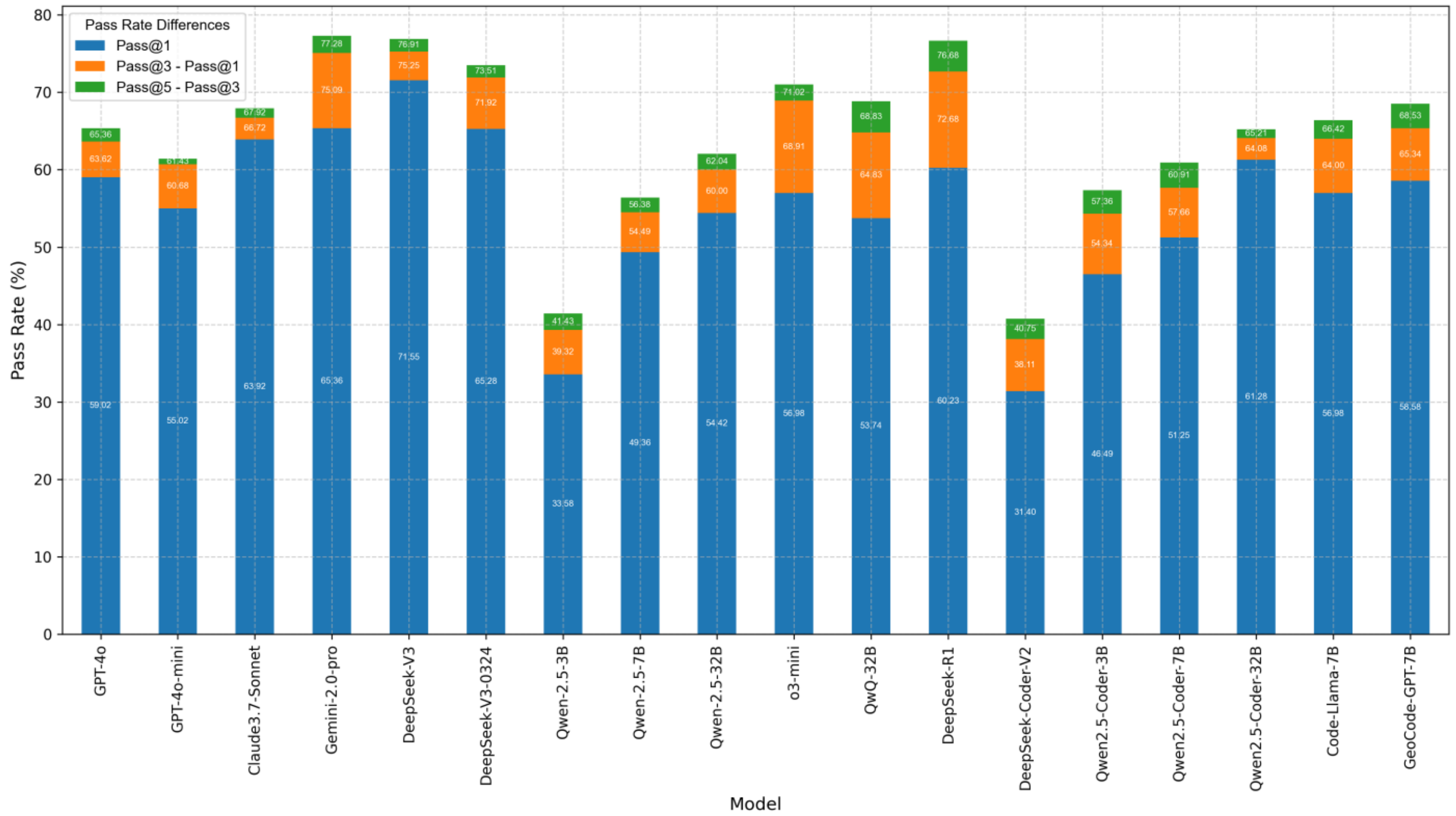

**Figure 8. Stacked bar chart of pass@n metrics.** The blue represents the pass@1 value, the orange represents the improvement of pass@3 over pass@1, and the green represents the improvement of pass@5 over pass@3. The white text on the bars indicates the absolute scores for pass@1, pass@3, and pass@5, respectively.

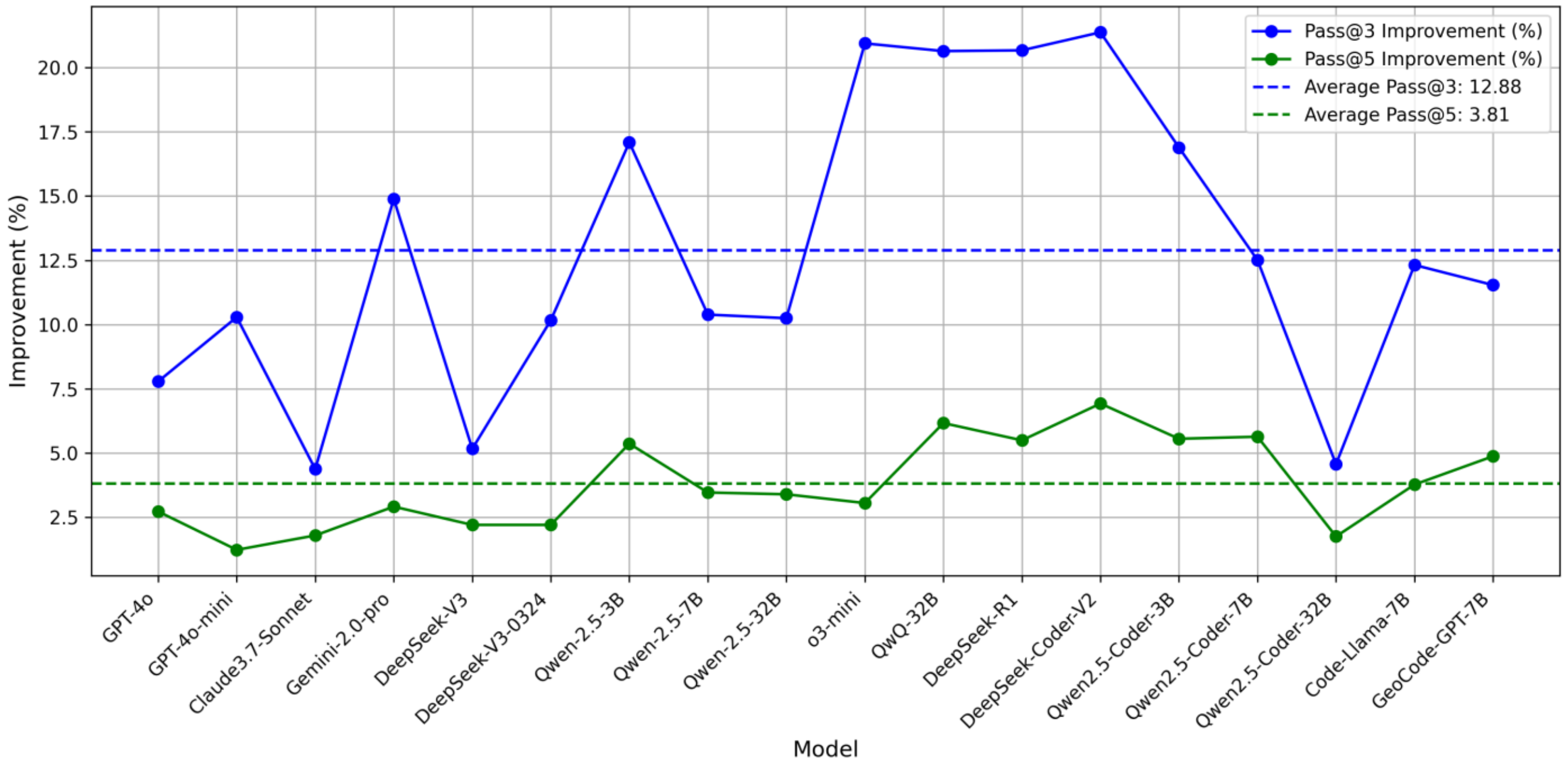

**Figure 9. Line chart of pass@3 and pass@5 improvement ratios.**

The bubble chart displaying the pass@n scores and relative rankings of all models is shown in Figure 10. Several key observations can be made:

- **Model performance ranking:** The dark blue bubbles, representing general-purpose non-reasoning models, generally occupy higher ranks, outperforming the red general-purpose reasoning models and pink general-purpose code generation models. The light blue bubble representing the geospatial code generation model GeoCode-GPT is positioned in the upper-middle tier, with an average rank of 7.33 among the 18 evaluated models.
- **Performance variation within the DeepSeek family:** DeepSeek-V3 (average rank 1.33), DeepSeek-V3-0324

(average rank 3.67), and DeepSeek-R1 (average rank 4.00) all rank among the top-performing models, demonstrating strong performance. However, DeepSeek-Coder-V2 performs poorly, ranking last (average rank 18.00), indicating that it lacks sufficient capability for GEE code generation tasks.

- **Inconsistent performance across model versions:** Surprisingly, DeepSeek-V3-0324, an optimized version of DeepSeek-V3, performs worse in GEE code generation, suggesting that later updates may not have specifically targeted improvements in this domain, potentially leading to performance degradation.
- **Performance of different parameter versions within the same model:** Significant differences are observed across parameter configurations of the same model. For instance, Qwen-2.5-Coder-32B (average rank 8.33) outperforms its 7B (rank 14.00) and 3B (rank 15.67) variants. Similarly, within the Qwen-2.5 family, the 32B version (rank 12.33) ranks notably higher than the 7B (rank 15.33) and 3B (rank 17.00) versions. In addition, GPT-4o (rank 9.33) also outperforms GPT-4o-mini (rank 12.00).
- **Performance gain of GeoCode-GPT-7B:** GeoCode-GPT-7B (average rank 7.33) outperforms its base model Code-Llama-7B (rank 9.50), indicating effective fine-tuning for GEE code generation tasks. However, the improvement is modest, possibly due to GeoCode-GPT's training covering a broad range of geospatial code types (e.g., ARCPY, GDAL), thus diluting its specialization in the GEE-specific domain.
- **Category-wise performance analysis:** Among the categories, the best-performing general-purpose non-reasoning LLM is DeepSeek-V3 (rank 1.33), the top general-purpose reasoning model is DeepSeek-R1 (rank 4.00), and the best general-purpose code generation model is Qwen-2.5-Coder-32B (rank 8.33).
- **Underwhelming performance of the GPT series:** The GPT series shows relatively weak performance. Specifically, GPT-4o (rank 9.33) and GPT-4o-mini (rank 12.00) are both outperformed by models from the DeepSeek, Claude, and Gemini families, as well as by GeoCode-GPT-7B. Even the GPT-series reasoning model o3-mini only marginally surpasses GeoCode-GPT-7B by less than one rank.

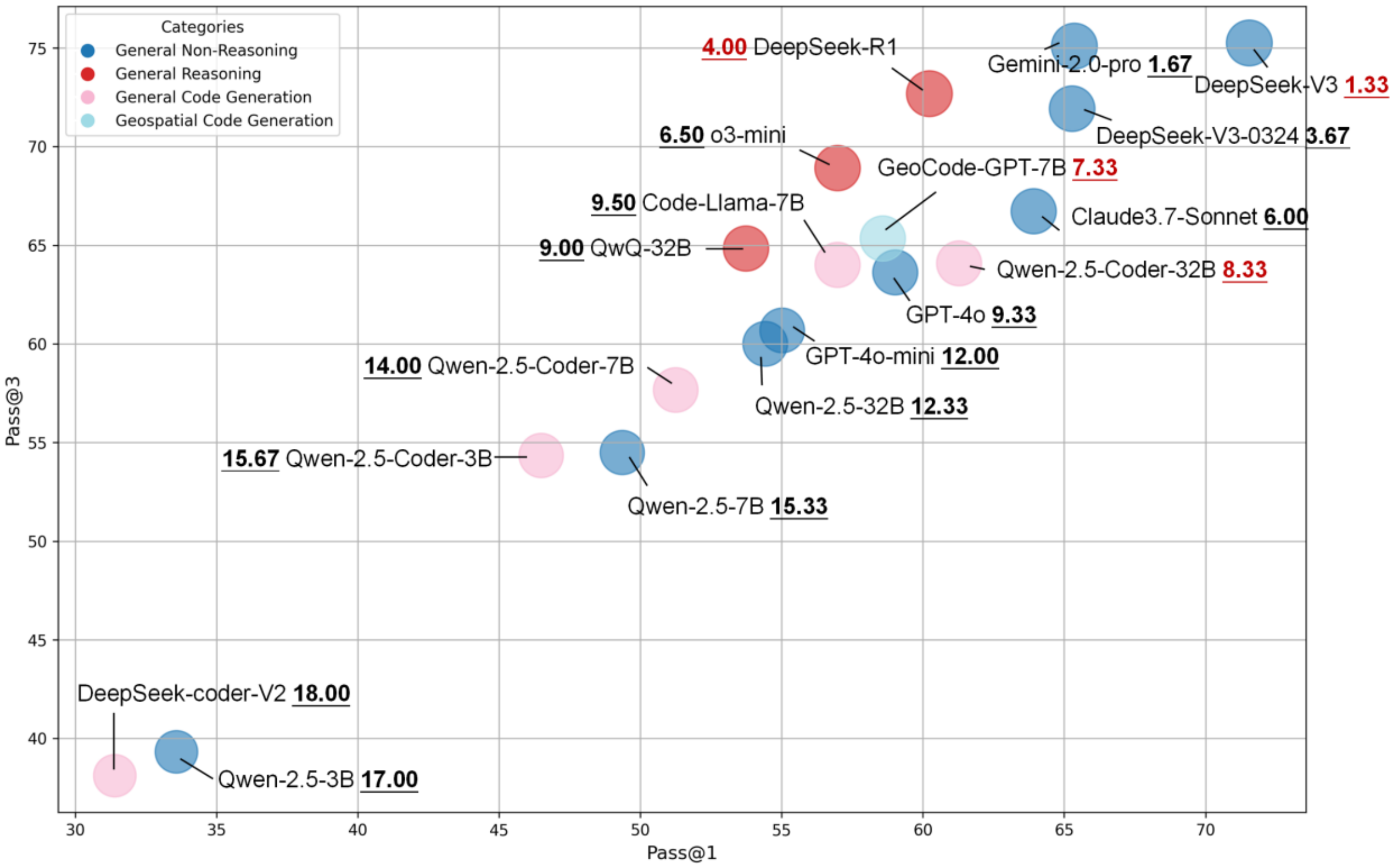

**Figure 10. LLM pass@n ranking bubble chart.** The x-axis represents the pass@1 scores, the y-axis represents the pass@3 scores, and the size of the bubbles corresponds to the pass@5 scores. Different colors represent different LLM types, as shown in the legend. The bold and underlined numbers beside the model names indicate the average ranking of the model under the pass@1, pass@3, and pass@5 metrics. The bold and underlined numbers in red represent the highest-ranking model within each LLM category.

To assess the stability of accuracy for the evaluated LLMs, we performed metric slicing and summarized the results in Table 7. Models with green shading indicate that both P_Rank and C_Rank are higher than S_Rank, suggesting that these models exhibit strong stability, with high overall rankings and robust consistency. Examples include DeepSeek-V3 and DeepSeek-V3-0324. Models with orange shading indicate that P_Rank is lower than both S_Rank and C_Rank. Although these models achieve high P_Rank, their poor stability leads to lower S_Rank scores. Typical examples include Gemini-2.0-pro, DeepSeek-R1, o3-mini, and QwQ-32B. Most of these are reasoning models, reflecting that poor stability is one of the current performance bottlenecks for reasoning-oriented LLMs. Models with blue shading indicate that P_Rank is higher than both S_Rank and C_Rank. Although P_Rank is not particularly high, these models demonstrate good stability and achieve relatively better rankings, making them more robust in scenarios where stability is crucial. Representative models include Claude3.7-Sonnet, Qwen2.5-Coder-32B, GPT-4o, GPT-4o-mini, and Qwen-2.5-7B.

**Table 7. Ranking of the models under pass@5, CV, and SA metrics.** P_Rank, C_Rank, and S_Rank represent the rankings based on pass@5, CV, and SA, respectively. Higher values of pass@5 and SA indicate better performance and higher ranking, while a lower CV value indicates better performance and a higher ranking. The table is sorted by S_Rank, reflecting the accuracy ranking of the models with the inclusion of stability factors, rather than solely considering accuracy. Category 1, 2, 3, and 4 correspond to General Non-Reasoning Models, General Reasoning Models, General Code Generation Models, and Geospatial Code Generation Models, respectively.

| Category | Model | pass@5 | CV | SA | P_Rank | C_Rank | S_Rank |
|---|---|---|---|---|---|---|---|
| 1 | DeepSeek-V3 | 76.91 | 0.07 | 71.9 | 2 | 3 | 1 |
| 1 | Gemini-2.0-pro | 77.28 | 0.154 | 66.95 | 1 | 11 | 2 |
| 1 | DeepSeek-V3-0324 | 73.51 | 0.112 | 66.11 | 4 | 6 | 3 |
| 1 | Claude3.7-Sonnet | 67.92 | 0.059 | 64.14 | 8 | 1 | 4 |
| 2 | DeepSeek-R1 | 76.68 | 0.215 | 63.14 | 3 | 16 | 5 |
| 3 | Qwen2.5-Coder-32B | 65.21 | 0.06 | 61.5 | 11 | 2 | 6 |
| 4 | GeoCode-GPT-7B | 68.53 | 0.145 | 59.84 | 7 | 10 | 7 |
| 1 | GPT-4o | 65.36 | 0.097 | 59.58 | 10 | 4 | 8 |
| 2 | o3-mini | 71.02 | 0.198 | 59.3 | 5 | 15 | 9 |
| 3 | Code-Llama-7B | 66.42 | 0.142 | 58.15 | 9 | 9 | 10 |
| 2 | QwQ-32B | 68.83 | 0.219 | 56.45 | 6 | 17 | 11 |
| 1 | GPT-4o-mini | 61.43 | 0.104 | 55.63 | 13 | 5 | 12 |
| 1 | Qwen-2.5-32B | 62.04 | 0.123 | 55.25 | 12 | 7 | 13 |
| 3 | Qwen2.5-Coder-7B | 60.91 | 0.159 | 52.57 | 14 | 12 | 14 |
| 1 | Qwen-2.5-7B | 56.38 | 0.125 | 50.14 | 16 | 8 | 15 |
| 3 | Qwen2.5-Coder-3B | 57.36 | 0.19 | 48.22 | 15 | 14 | 16 |
| 1 | Qwen-2.5-3B | 41.43 | 0.189 | 34.83 | 17 | 13 | 17 |
| 3 | DeepSeek-Coder-V2 | 40.75 | 0.229 | 33.14 | 18 | 18 | 18 |

### 6.2. Resource consumption

The evaluation results for resource consumption are presented in Table 8. This study provides visual analyses of token consumption, inference time, and the number of core generated code lines.

**Table 8. Evaluation results for resource consumption.** For the QwQ-32B model using API calls, due to the provider's configuration, only "streaming calls" are supported. In this mode, Token consumption cannot be tracked, so it is marked as N/A.

| Model | Inference Method | Tok. (tokens) | In.T (s) | Co.L (lines) |
|---|---|---|---|---|
| General Non-Reasoning | | | | |
| GPT-4o | API call | 210 | 3.31 | 7.77 |

| Model | Inference Method | Tok. (tokens) | In.T (s) | Co.L (lines) |
|---|---|---|---|---|
| GPT-4o-mini | API call | 208 | 7.63 | 5.86 |
| Claude3.7-Sonnet | API call | 265 | 11.72 | 8.98 |
| Gemini-2.0-pro | API call | 223 | 24.55 | 5.2 |
| DeepSeek-V3 | API call | 190 | 8.87 | 4.86 |
| DeepSeek-V3-0324 | API call | 204 | 16.32 | 6.82 |
| Qwen-2.5-3B | Local deployment | 186 | 2.58 | 4.12 |
| Qwen-2.5-7B | Local deployment | 197 | 3.88 | 6.28 |
| Qwen-2.5-32B | API call | 205 | 5.63 | 6.6 |
| **General Reasoning** | | | | |
| o3-mini | API call | 1083 | 7.40 | 6.93 |
| QwQ-32B | API call | N/A | 44.68 | 5.64 |
| DeepSeek-R1 | API call | 1557 | 78.30 | 5.32 |
| **General Code Generation** | | | | |
| DeepSeek-Coder-V2 | Local deployment | 285 | 8.39 | 10.06 |
| Qwen2.5-Coder-3B | Local deployment | 240 | 2.51 | 9.11 |
| Qwen2.5-Coder-7B | Local deployment | 224 | 3.76 | 7.06 |
| Qwen2.5-Coder-32B | API call | 198 | 5.50 | 5.79 |
| Code-Llama-7B | Local deployment | 256 | 3.05 | 3.58 |
| **Geospatial Code Generation** | | | | |
| GeoCode-GPT-7B | Local deployment | 253 | 4.05 | 11.79 |

The bar chart of average token consumption for GEE code generation across all LLMs is shown in Figure 11. The results show that the General Non-Reasoning, General Code Generation, and Geospatial Code Generation model categories exhibit relatively similar levels of token consumption, while the General Reasoning models consume significantly more tokens—approximately 6 to 7 times higher on average than the other three categories. This finding provides a useful reference for users in estimating token-based billing costs when selecting a model. It suggests that, for the same GEE code generation task, General Reasoning models will incur 6 to 7 times the cost compared to General Non-Reasoning, General Code Generation, and Geospatial Code Generation models.

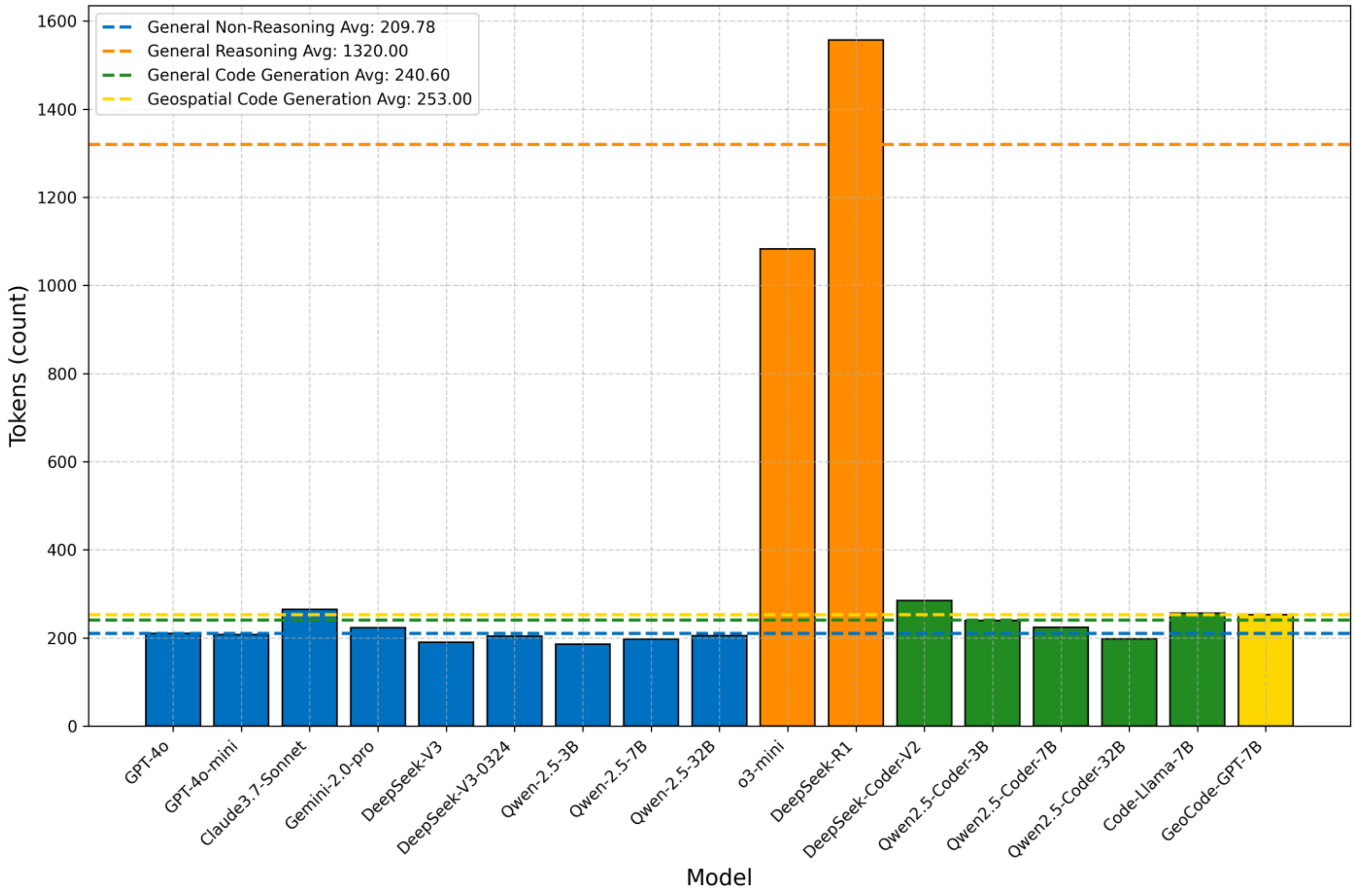

**Figure 11. Average token consumption across LLMs.**

The lollipop chart of inference time consumption for GEE code generation across LLMs is shown in Figure 12. In terms of inference methods, models using the API call approach (circles) exhibit longer inference times compared to those using local deployment (squares). This may be due to network latency and limitations in the computing resources of remote servers. From a model category perspective, general reasoning models (orange) generally require more inference time than other types. However, o3-mini is an exception—its inference latency is even lower than that of the locally deployed DeepSeek-Coder-V2, indicating that its server-side computational resources may have been optimized accordingly. In addition, the average inference time per unit test case for DeepSeek-R1 and QwQ-32B reaches as high as 78.3 seconds and 44.68 seconds, respectively—2 to 40 times longer than other models—indicating that these two models are in urgent need of targeted optimization for inference latency.

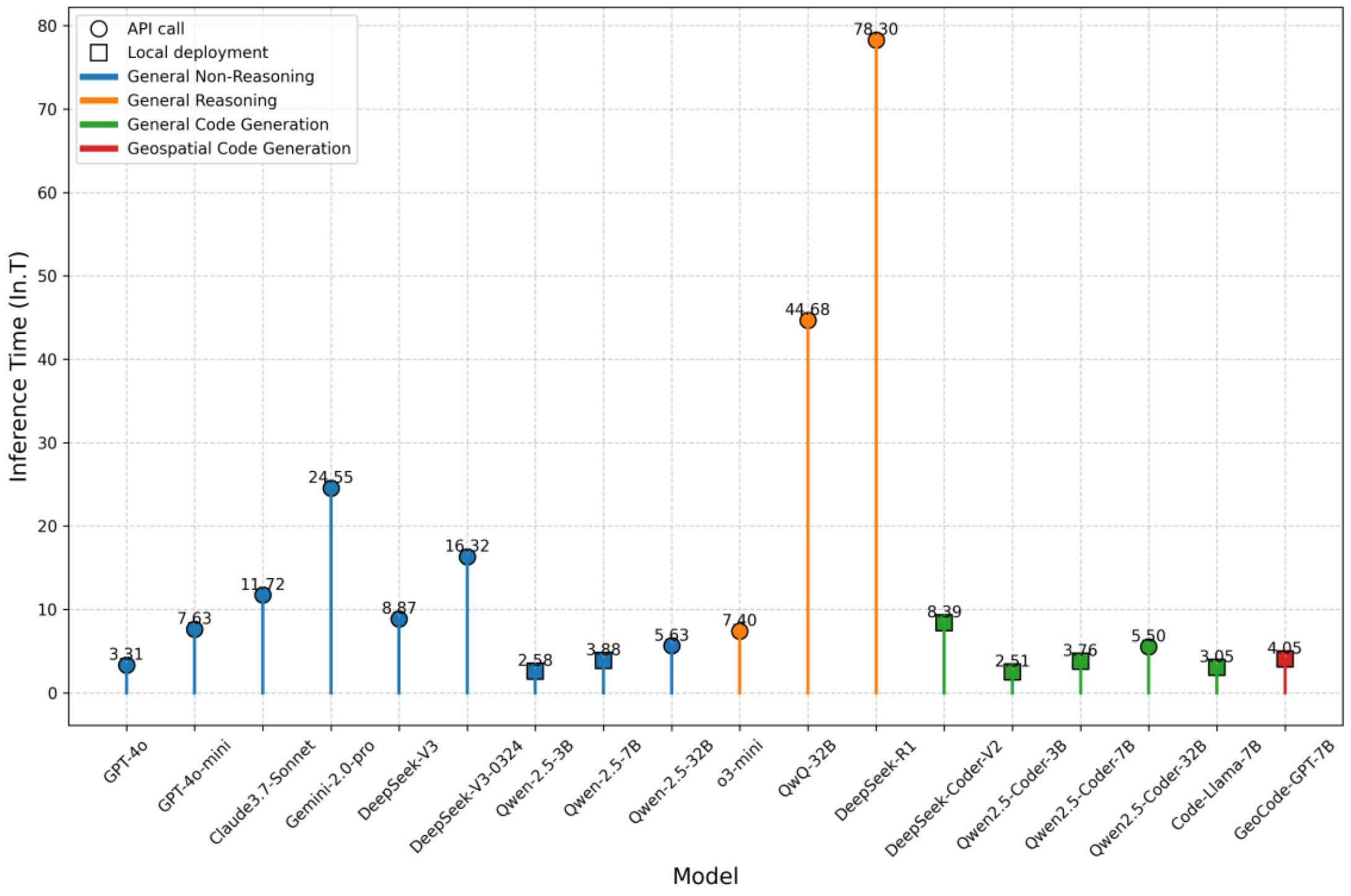

**Figure 12. Average inference time comparison of LLMs**

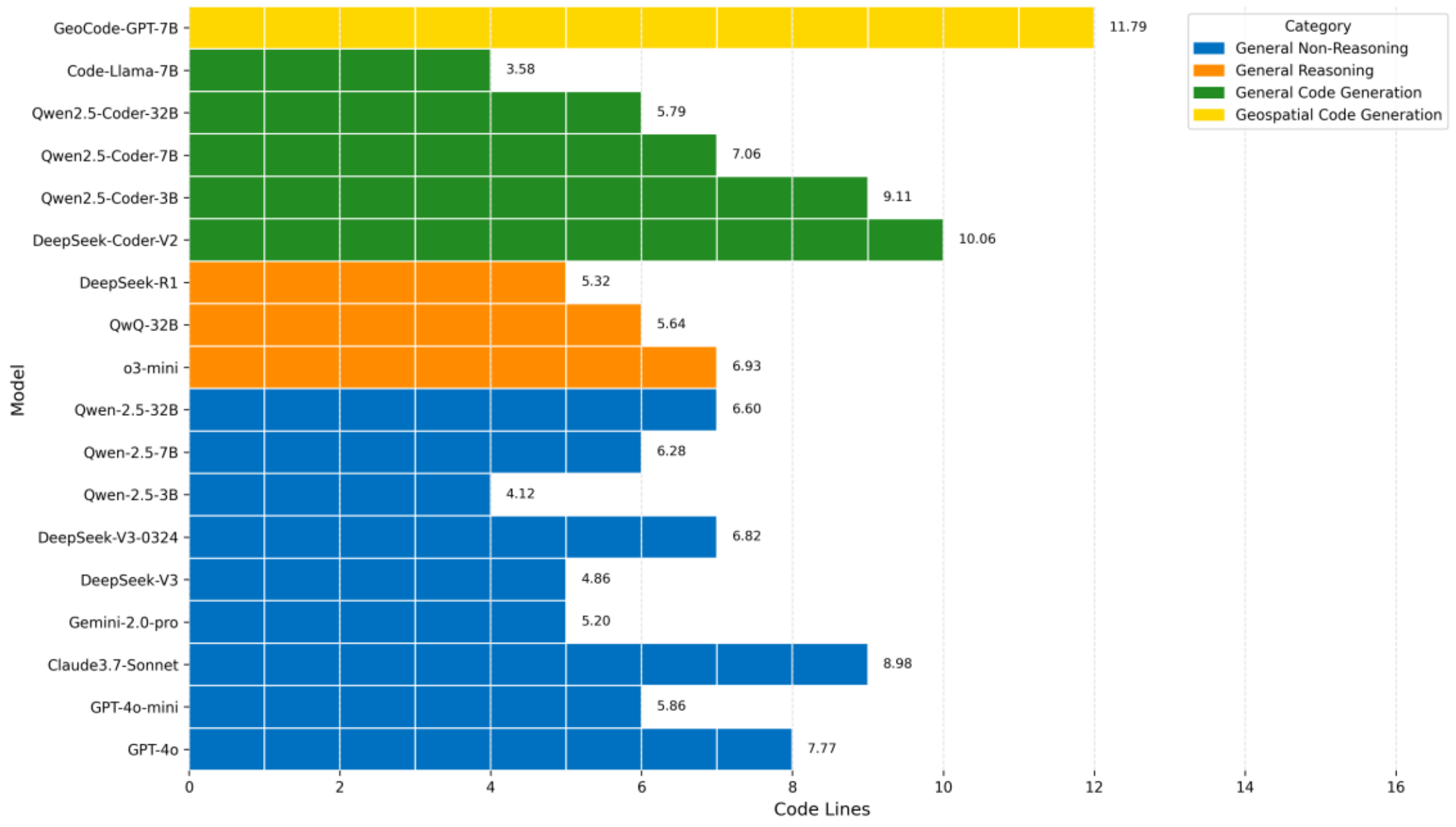

**Figure 13. Average lines of generated GEE code per model**

The token consumption metric reflects not only the length of the generated code but also includes the model's reasoning output and the length of the prompt template, thereby representing more of the reasoning cost than the actual size of the generated code itself. To more accurately measure the structural length of the model's output code, we excluded the influence of prompt and reasoning-related content and used the total number of generated lines of code (including both comments and executable lines) as the evaluation metric. The results are shown in Figure 13. As observed, GeoCode-GPT-7B (average: 11.79 lines), DeepSeek-Coder-V2 (10.06), Qwen2.5-Coder-3B (9.11), and Claude3.7-Sonnet (8.98) rank among the highest in terms of code length. This may be attributed to excessive generated comments or more standardized code structures that automatically include formal comment templates, thereby increasing the overall line count. Additionally, a noteworthy phenomenon is observed within the Qwen2.5-Coder family: models with larger parameter sizes tend to generate shorter code. For example, the Qwen2.5-Coder-32B model has an average code length of 5.79 lines, which is significantly shorter than its 7B (7.06) and 3B (9.11) versions. This result contradicts conventional expectations and may suggest that larger models possess stronger capabilities in code compression and refinement, or that their output formatting is subject to stricter constraints and optimizations during training.

### 6.3. Operational efficiency

The operational efficiency results for each model are presented in Table 9.

**Table 9. Evaluation results for operational efficiency**

| Model | Inference Method | Tok.-E | In.T-E | Co.L-E |
|---|---|---|---|---|
| **General Non-Reasoning** | | | | |
| GPT-4o | API call | 0.311 | 19.746 | 7.77 |
| GPT-4o-mini | API call | 0.295 | 8.052 | 5.86 |
| Claude3.7-Sonnet | API call | 0.256 | 5.796 | 8.98 |
| Gemini-2.0-pro | API call | 0.347 | 3.148 | 5.2 |
| DeepSeek-V3 | API call | 0.405 | 8.670 | 4.86 |
| DeepSeek-V3-0324 | API call | 0.360 | 4.504 | 6.82 |
| Qwen-2.5-3B | Local deployment | 0.223 | 16.060 | 4.12 |
| Qwen-2.5-7B | Local deployment | 0.286 | 14.530 | 6.28 |
| Qwen-2.5-32B | API call | 0.303 | 11.019 | 6.6 |
| **General Reasoning** | | | | |
| o3-mini | API call | 0.066 | 9.597 | 6.93 |
| QwQ-32B | API call | N/A | 1.541 | 5.64 |
| DeepSeek-R1 | API call | 0.049 | 0.979 | 5.32 |
| **General Code Generation** | | | | |
| DeepSeek-Coder-V2 | Local deployment | 0.143 | 8.39 | 10.06 |
| Qwen2.5-Coder-3B | Local deployment | 0.239 | 2.51 | 9.11 |
| Qwen2.5-Coder-7B | Local deployment | 0.272 | 3.76 | 7.06 |
| Qwen2.5-Coder-32B | API call | 0.329 | 5.50 | 5.79 |
| Code-Llama-7B | Local deployment | 0.259 | 3.05 | 3.58 |
| **Geospatial Code Generation** | | | | |
| GeoCode-GPT-7B | Local deployment | 0.271 | 4.05 | 11.79 |

Since efficiency metrics are expressed as ratios between numerators and denominators, they lack direct interpretability or intuitive real-world meaning. To address this, we independently ranked the three efficiency metrics—Token Efficiency (Tok.-E), Inference Time Efficiency (In.T-E), and Code Line Efficiency (Co.L-E)—denoted as T_Rank, I_Rank, and Co_Rank, respectively. Their average was then calculated to produce an overall efficiency ranking, denoted as E_Rank. In addition, we cross-referenced P_Rank (ranking based solely on accuracy, from Table 7) and S_Rank (ranking based on stability-adjusted accuracy) to perform a comparative slicing of the ranking indicators. Based on this analysis, we constructed Table 10 to provide a comprehensive evaluation of each model's overall performance. It is important to note that P_Rank reflects pure accuracy ranking, E_Rank captures performance when efficiency factors are considered, and S_Rank reflects accuracy adjusted for stability. Finally, we define the average of these three rankings as the Total Performance Ranking, denoted as Total_Rank. According to the results, DeepSeek-V3, Gemini-2.0-pro, and DeepSeek-V3-0324 consistently rank at the top across all three dimensions and demonstrate excellent overall performance. All three are commercial models, making them suitable for API-based deployment. In contrast, models such as Code-Llama-7B, Qwen2.5-Coder-32B, and GPT-4o do not rank as highly in terms of P_Rank and S_Rank, but their strong performance in E_Rank makes them well-suited for local deployment (the first two) or for scenarios requiring high generation efficiency (GPT-4o). By comparison, although models like DeepSeek-R1, GeoCode-GPT-7B, o3-mini, and Claude3.7-Sonnet perform well in terms of accuracy and stability, their low E_Rank scores lead to less favorable overall rankings, indicating a need to improve generation efficiency in order to optimize their total performance.

**Table 10. Rank-based comparative evaluation of models.** The table is sorted by Total_Rank in ascending order. If models share the same average rank, they are assigned the same ranking (e.g., both DeepSeek-V3 and Code-Llama-7B are ranked 1 in E_Rank). Blue highlights indicate the top 12 models in E_Rank, green for the top 12 in S_Rank, and orange for the top 12 in P_Rank. Gray highlights mark the bottom 6 models across E_Rank, S_Rank, and P_Rank. Categories 1, 2, 3, and 4 correspond to General Non-Reasoning Models, General Reasoning Models, General Code Generation Models, and Geospatial Code Generation Models, respectively.

| Category | Model | T_Rank | I_Rank | Co_Rank | E_Rank | S_Rank | P_Rank | Total_Rank |
|---|---|---|---|---|---|---|---|---|
| 1 | DeepSeek-V3 | 1 | 11 | 2 | 1 | 1 | 2 | **1** |
| 1 | Gemini-2.0-pro | 3 | 16 | 3 | 4 | 2 | 1 | **2** |
| 1 | DeepSeek-V3-0324 | 2 | 15 | 7 | 6 | 3 | 4 | **3** |
| 3 | Code-Llama-7B | 11 | 2 | 1 | 1 | 10 | 9 | **4** |
| 3 | Qwen2.5-Coder-32B | 4 | 8 | 6 | 3 | 6 | 11 | **4** |
| 1 | GPT-4o | 5 | 3 | 14 | 5 | 8 | 10 | **6** |
| 2 | DeepSeek-R1 | 17 | 18 | 4 | 15 | 5 | 3 | **6** |
| 4 | GeoCode-GPT-7B | 10 | 4 | 17 | 13 | 7 | 7 | **8** |
| 2 | o3-mini | 16 | 10 | 9 | 14 | 9 | 5 | **9** |
| 1 | Claude3.7-Sonnet | 12 | 13 | 15 | 17 | 4 | 8 | **10** |
| 1 | Qwen-2.5-32B | 6 | 9 | 11 | 7 | 13 | 12 | **11** |
| 1 | GPT-4o-mini | 7 | 12 | 8 | 8 | 12 | 13 | **12** |
| 2 | QwQ-32B | 18 | 17 | 5 | 16 | 11 | 6 | **12** |
| 3 | Qwen2.5-Coder-7B | 9 | 5 | 13 | 9 | 14 | 14 | **14** |
| 1 | Qwen-2.5-7B | 8 | 7 | 12 | 10 | 15 | 16 | **15** |
| 3 | Qwen2.5-Coder-3B | 13 | 1 | 16 | 11 | 16 | 15 | **16** |
| 1 | Qwen-2.5-3B | 14 | 6 | 10 | 12 | 17 | 17 | **17** |
| 3 | DeepSeek-Coder-V2 | 15 | 14 | 18 | 18 | 18 | 18 | **18** |

## 6.4. Error type logs

The types of errors encountered by each model during GEE code generation are summarized in Table 11, revealing an overall consistent error pattern across models. Parameter errors occur at a significantly higher rate than invalid answers, while syntax errors and network errors appear only sporadically and at extremely low frequencies. This suggests that the core challenge currently faced by models in GEE-based geospatial code generation lies in the lack of domain-specific parameter knowledge, including references to platform-integrated datasets, band names, coordinate formats, and other geoscientific details. As such, there is an urgent need to augment training data with domain-relevant knowledge specific to the GEE platform and to implement targeted fine-tuning. Meanwhile, the models have demonstrated strong stability in terms of basic syntax, code structure, and loop control, with related errors being extremely rare. This indicates that their foundational programming capabilities are largely mature. Therefore, future optimization efforts should shift toward enhancing domain knowledge, rather than further reinforcing general coding skills.

Table 11. Error Type Distribution in GEE Code Generation Across Models

| Model | Parameter Error (%) | Invalid Answer (%) | Syntax Error (%) | Network Error (%) |
|---|---|---|---|---|
| **General Non-Reasoning** | | | | |
| GPT-4o | 72.21 | 26.58 | 1.02 | 0.19 |
| GPT-4o-mini | 75.88 | 22.29 | 1.49 | 0.34 |
| Claude3.7-Sonnet | 65.81 | 31.92 | 1.76 | 0.50 |
| Gemini-2.0-pro | 55.71 | 37.15 | 7.01 | 0.13 |
| DeepSeek-V3 | 72.75 | 26.29 | 0.37 | 0.58 |
| DeepSeek-V3-0324 | 79.40 | 19.86 | 0.43 | 0.30 |
| Qwen-2.5-3B | 83.72 | 8.38 | 7.90 | 0.00 |
| Qwen-2.5-7B | 83.44 | 12.60 | 3.96 | 0.00 |
| Qwen-2.5-32B | 78.47 | 18.65 | 2.75 | 0.13 |
| **General Reasoning** | | | | |
| o3-mini | 67.79 | 30.02 | 1.84 | 0.35 |
| QwQ-32B | 85.68 | 13.01 | 1.11 | 0.21 |
| DeepSeek-R1 | 85.04 | 14.62 | 0.19 | 0.15 |
| **General Code Generation** | | | | |
| DeepSeek-Coder-V2 | 84.47 | 10.62 | 4.78 | 0.13 |
| Qwen2.5-Coder-3B | 75.26 | 12.54 | 12.20 | 0.00 |
| Qwen2.5-Coder-7B | 84.76 | 14.42 | 0.63 | 0.19 |
| Qwen2.5-Coder-32B | 79.19 | 19.96 | 0.43 | 0.43 |
| Code-Llama-7B | 80.01 | 18.47 | 1.37 | 0.14 |
| **Geospatial Code Generation** | | | | |
| GeoCode-GPT-7B | 77.21 | 9.54 | 13.14 | 0.11 |

### 6.5. Key findings and insights

Based on the AutoGEEval evaluation framework, this study systematically assessed the overall performance of 18 LLMs in geospatial code generation tasks across four dimensions: accuracy, resource consumption, operational efficiency, and error types. The main findings are as follows:

- **In terms of accuracy**, multiple rounds of generation help alleviate hallucination and improve output stability. However, a diminishing marginal return is observed—pass@3 shows a significantly larger improvement than pass@5, indicating that model optimization should focus on improving the first few generations. Based on the CV and SA metrics, DeepSeek-V3 and Gemini-2.0-pro achieve a good balance between accuracy and stability.
- **In terms of resource consumption**, general-purpose reasoning models (e.g., DeepSeek-R1, QwQ-32B) consume significantly more tokens and inference time compared to other models, resulting in high computational cost and slow response, with average generation times 2 to 40 times longer than other models. These models urgently require latency optimization. In contrast, general non-reasoning models and code generation models offer better cost-performance and are more suitable for high-frequency usage scenarios.
- **In terms of operational efficiency**, considering tokens, time, and code structure, DeepSeek-V3, Gemini-2.0-pro, and Code-Llama-7B stand out, showing significant cost-effectiveness. Some models (e.g., GeoCode-GPT-7B) achieve acceptable accuracy but suffer from low efficiency, limiting their practical applicability in production environments.
- **Error type analysis reveals that parameter errors are the most frequent, while syntax and network errors are rare.** This indicates that most models have already achieved mature capabilities in basic syntax and code execution. However, the lack of domain-specific knowledge required by the GEE platform (e.g., dataset paths, band names, coordinate formats) remains a major weakness, highlighting the need for targeted fine-tuning using domain-specific data.
- **Performance varies significantly across models.** The DeepSeek family performs exceptionally well overall, with DeepSeek-V3 ranking first across multiple metrics, demonstrating excellent stability and generalizability. In contrast, DeepSeek-Coder-V2 ranks lowest, revealing adaptability differences even within the same model family.

GeoCode-GPT shows modest improvement over its base model but lacks clear advantages on GEE tasks, suggesting the need for more focused training. The GPT series delivers average performance, clearly outperformed by the DeepSeek, Claude, and Gemini families.

- **Model size is not the decisive factor in performance.** Several cases demonstrate that "bigger is not always better": for instance, Qwen2.5-Coder-32B outperforms its 7B and 3B versions in accuracy and efficiency, but its code structure is less concise, and its stability is inferior to some smaller models (e.g., Claude3.7-Sonnet). This suggests that performance in specific tasks depends more on fine-tuning quality, instruction alignment, and output formatting than on model size alone.
- Finally, **model selection** should be guided by the overall ranking indicator (Total_Rank), which integrates accuracy (P_Rank), stability (S_Rank), and efficiency (E_Rank). From this perspective, models with high accuracy and high efficiency, such as DeepSeek-V3, are ideal for high-performance, high-frequency production API deployment; those with high accuracy and strong stability, like Claude3.7-Sonnet, are better suited for scientific and engineering tasks requiring output consistency; models offering high efficiency and support for local deployment, such as Code-Llama-7B and Qwen2.5-Coder-32B, are appropriate for edge computing or cost-sensitive batch generation scenarios; whereas reasoning models like DeepSeek-R1 and QwQ-32B, despite their accuracy, are less suitable for latency- or cost-constrained applications due to their low efficiency and limited stability.

## 7. Conclusion

This study presents AutoGEEval, the first automated evaluation framework designed for geospatial code generation tasks on the GEE platform. Implemented via the Python API, the framework supports unit-level, multimodal, and end-to-end evaluation across 26 GEE data types. It consists of three core components: the constructed benchmark (AutoGEEval-Bench) with 1325 unit test cases; the Submission Program, which guides LLMs to generate executable code via prompts; and the Judge Program, which automatically verifies output correctness, resource consumption, and error types. Using this framework, we conducted a comprehensive evaluation of 18 representative LLMs, spanning general-purpose, reasoning-enhanced, code generation, and geospatial-specialized models. The results reveal performance gaps, code hallucination phenomena, and trade-offs in code quality, offering valuable insights for the optimization of future geospatial code generation technologies.

### 7.1. Significance and contributions

This study is the first to establish a dedicated evaluation system for LLMs in geospatial code generation tasks, addressing key gaps in current tools that lack geospatial coverage, granularity, and automation. Through the proposed AutoGEEval framework, we achieved systematic evaluation of multimodal GEE data types and API function call capabilities, advancing the automated transformation from natural language to geospatial code. Compared to existing methods that rely heavily on manual scoring, AutoGEEval offers high automation, standardization, and reproducibility, substantially reducing evaluation costs and improving efficiency. The framework supports comprehensive tracking and quantitative analysis of code correctness, inference efficiency, resource consumption, and error types, providing a clear indicator system and real-world entry points for model refinement. Moreover, the constructed benchmark AutoGEEval-Bench, covering 1325 test cases and 26 GEE data types, is both scalable and representative, serving as a valuable public resource for future research on intelligent geospatial code generation. Overall, this work advances the transformation of geospatial code generation from an engineering tool into a quantifiable scientific problem, and provides a methodological reference and practical blueprint for interdisciplinary AI model evaluation paradigms.

### 7.2. Limitations and future work

Despite the representativeness of the proposed unit-level evaluation framework, several limitations remain, and future work can explore multiple directions for further enhancement. Currently, the evaluation tasks focus on single-function unit tests, and although 1325 use cases are included, coverage remains limited. Future expansions could include a broader test set, especially under boundary conditions and abnormal inputs, to evaluate model robustness under extreme scenarios. Additionally, introducing function composition and cross-API test cases will allow assessment of model capabilities in handling complex logical structures. The current 26 GEE data types could also be expanded using modality-based classification strategies to achieve a more balanced and comprehensive benchmark. In terms of evaluation metrics, the current system primarily centers on execution correctness. Future extensions could incorporate multi-dimensional evaluation criteria, including code structural complexity, runtime efficiency, and resource usage. Given the continual evolution of LLMs, a valuable next step would be to build an open, continuous evaluation platform that includes economic cost dimensions and releases "cost-effectiveness leaderboards", thereby driving community development and enhancing the visibility and influence of geospatial code generation research.

**ORCID**

Shuyang Hou: https://orcid.org/0009-0000-6984-9959
Huayi Wu: https://orcid.org/0000-0003-3971-0512

**Data availability statement**

The experimental data used in this study can be downloaded from https://github.com/szx-0633/AutoGEEval. Other data that support the findings of this study are available from the corresponding author upon reasonable request.

**Funding**

The work was supported by the National Natural Science Foundation of China [Grant number 41971349]

**References**

1. Li, Y.; Choi, D.; Chung, J.; Kushman, N.; Schrittwieser, J.; Leblond, R.; Eccles, T.; Keeling, J.; Gimeno, F.; Dal Lago, A. Competition-level code generation with alphacode. *Science* **2022**, *378*, 1092-1097.
2. Popat, S.; Starkey, L. Learning to code or coding to learn? A systematic review. *Computers & Education* **2019**, *128*, 365-376.
3. Bonner, A.J.; Kifer, M. An overview of transaction logic. *Theoretical Computer Science* **1994**, *133*, 205-265.
4. Jiang, J.; Wang, F.; Shen, J.; Kim, S.; Kim, S. A survey on large language models for code generation. *arXiv preprint arXiv:2406.00515* **2024**.
5. Wang, J.; Chen, Y. A review on code generation with llms: Application and evaluation. 2023; pp. 284-289.
6. Guo, D.; Zhu, Q.; Yang, D.; Xie, Z.; Dong, K.; Zhang, W.; Chen, G.; Bi, X.; Wu, Y.; Li, Y.K. DeepSeek-Coder: When the Large Language Model Meets Programming--The Rise of Code Intelligence. *arXiv preprint arXiv:2401.14196* **2024**.
7. Hui, B.; Yang, J.; Cui, Z.; Yang, J.; Liu, D.; Zhang, L.; Liu, T.; Zhang, J.; Yu, B.; Lu, K. Qwen2. 5-coder technical report. *arXiv preprint arXiv:2409.12186* **2024**.
8. Roziere, B.; Gehring, J.; Gloeckle, F.; Sootla, S.; Gat, I.; Tan, X.E.; Adi, Y.; Liu, J.; Sauvestre, R.; Remez, T. Code llama: Open foundation models for code. *arXiv preprint arXiv:2308.12950* **2023**.
9. Rahman, M.M.; Kundu, A. Code Hallucination. *arXiv preprint arXiv:2407.04831* **2024**.
10. Li, D.; Murr, L. HumanEval on Latest GPT Models--2024. *arXiv preprint arXiv:2402.14852* **2024**.
11. Yu, Z.; Zhao, Y.; Cohan, A.; Zhang, X.-P. HumanEval Pro and MBPP Pro: Evaluating Large Language Models on Self-invoking Code Generation. *arXiv preprint arXiv:2412.21199* **2024**.
12. Jain, N.; Han, K.; Gu, A.; Li, W.-D.; Yan, F.; Zhang, T.; Wang, S.; Solar-Lezama, A.; Sen, K.; Stoica, I. Livecodebench: Holistic and contamination free evaluation of large language models for code. *arXiv preprint arXiv:2403.07974* **2024**.
13. Gentleman, R.C.; Carey, V.J.; Bates, D.M.; Bolstad, B.; Dettling, M.; Dudoit, S.; Ellis, B.; Gautier, L.; Ge, Y.; Gentry, J. Bioconductor: open software development for computational biology and bioinformatics. *Genome biology* **2004**, *5*, 1-16.
14. Varma, J.R.; Virmani, V. Computational finance using QuantLib-Python. *Computing in Science & Engineering* **2016**, *18*, 78-88.

15. Huber, W.; Carey, V.J.; Gentleman, R.; Anders, S.; Carlson, M.; Carvalho, B.S.; Bravo, H.C.; Davis, S.; Gatto, L.; Girke, T. Orchestrating high-throughput genomic analysis with Bioconductor. *Nature methods* **2015**, *12*, 115-121.
16. Amezquita, R.A.; Lun, A.T.L.; Becht, E.; Carey, V.J.; Carpp, L.N.; Geistlinger, L.; Marini, F.; Rue-Albrecht, K.; Risso, D.; Soneson, C. Orchestrating single-cell analysis with Bioconductor. *Nature methods* **2020**, *17*, 137-145.
17. Shen, L.; Chen, X.; Liu, R.; Wang, H.; Ji, G. Domain-specific language techniques for visual computing: a comprehensive study. *Archives of Computational Methods in Engineering* **2021**, *28*, 3113-3134.
18. Gu, X.; Chen, M.; Lin, Y.; Hu, Y.; Zhang, H.; Wan, C.; Wei, Z.; Xu, Y.; Wang, J. On the effectiveness of large language models in domain-specific code generation. *ACM Transactions on Software Engineering and Methodology* **2025**, *34*, 1-22.
19. Capolupo, A.; Monterisi, C.; Caporusso, G.; Tarantino, E. Extracting land cover data using GEE: A review of the classification indices. 2020; pp. 782-796.
20. Tamiminia, H.; Salehi, B.; Mahdianpari, M.; Quackenbush, L.; Adeli, S.; Brisco, B. Google Earth Engine for geo-big data applications: A meta-analysis and systematic review. *ISPRS journal of photogrammetry and remote sensing* **2020**, *164*, 152-170.
21. Ratti, C.; Wang, Y.; Ishii, H.; Piper, B.; Frenchman, D. Tangible User Interfaces (TUIs): a novel paradigm for GIS. *Transactions in GIS* **2004**, *8*, 407-421.
22. Zhao, Q.; Yu, L.; Li, X.; Peng, D.; Zhang, Y.; Gong, P. Progress and trends in the application of Google Earth and Google Earth Engine. *Remote Sens.* **2021**, *13*, 3778.
23. Mutanga, O.; Kumar, L. Google earth engine applications. **2019**, *11*, 591.
24. Hou, S.; Shen, Z.; Zhao, A.; Liang, J.; Gui, Z.; Guan, X.; Li, R.; Wu, H. GeoCode-GPT: A large language model for geospatial code generation. *International Journal of Applied Earth Observation and Geoinformation* **2025**, 104456.
25. Hou, S.; Liang, J.; Zhao, A.; Wu, H. GEE-OPs: An Operator Knowledge Base for Geospatial Code Generation on the Google Earth Engine Platform Powered by Large Language Models. *arXiv preprint arXiv:2412.05587* **2024**.
26. Yang, L.; Driscol, J.; Sarigai, S.; Wu, Q.; Chen, H.; Lippitt, C.D. Google Earth Engine and artificial intelligence (AI): a comprehensive review. *Remote Sens.* **2022**, *14*, 3253.
27. Hou, S.; Shen, Z.; Liang, J.; Zhao, A.; Gui, Z.; Li, R.; Wu, H. Can large language models generate geospatial code? *arXiv preprint arXiv:2410.09738* **2024**.
28. Gramacki, P.; Martins, B.; Szymański, P. Evaluation of Code LLMs on Geospatial Code Generation. *arXiv preprint arXiv:2410.04617* **2024**.
29. Hou, S.; Jiao, H.; Shen, Z.; Liang, J.; Zhao, A.; Zhang, X.; Wang, J.; Wu, H. Chain-of-Programming (CoP): Empowering Large Language Models for Geospatial Code Generation. *arXiv preprint arXiv:2411.10753* **2024**.
30. Hou, S.; Shen, Z.; Zhao, A.; Liang, J.; Gui, Z.; Guan, X.; Li, R.; Wu, H. GeoCode-GPT: A Large Language Model for Geospatial Code Generation Tasks. *arXiv preprint arXiv:2410.17031* **2024**.
31. Hou, S.; Zhao, A.; Liang, J.; Shen, Z.; Wu, H. Geo-FuB: A Method for Constructing an Operator-Function Knowledge Base for Geospatial Code Generation Tasks Using Large Language Models. *arXiv preprint arXiv:2410.20975* **2024**.
32. Yang, C.; Huang, Q.; Li, Z.; Liu, K.; Hu, F. Big Data and cloud computing: innovation opportunities and challenges. *International Journal of Digital Earth* **2017**, *10*, 13-53.
33. Xu, L.; Zhao, S.; Lin, Q.; Chen, L.; Luo, Q.; Wu, S.; Ye, X.; Feng, H.; Du, Z. Evaluating large language models on geospatial tasks: a multiple geospatial task benchmarking study. *International Journal of Digital Earth* **2025**, *18*, 2480268.
34. Janowicz, K.; Gao, S.; McKenzie, G.; Hu, Y.; Bhaduri, B. GeoAI: spatially explicit artificial intelligence techniques for geographic knowledge discovery and beyond. **2020**, *34*, 625-636.
35. Fisher, T.; MacDonald, C. *An overview of the Canada geographic information system (CGIS)*; Lands Directorate Environment Canada Ottawa, ON, Canada: 1980.
36. Neteler, M.; Bowman, M.H.; Landa, M.; Metz, M. GRASS GIS: A multi-purpose open source GIS. *Environmental Modelling & Software* **2012**, *31*, 124-130.
37. Zhang, M.; Yue, P.; Guo, X. GIScript: Towards an interoperable geospatial scripting language for GIS programming. 2014; pp. 1-5.
38. Coetzee, S.; Ivánová, I.; Mitasova, H.; Brovelli, M.A. Open geospatial software and data: A review of the current state and a perspective into the future. *ISPRS Int. J. Geo-Inf.* **2020**, *9*, 90.
39. White, P.; Powell, S. Code-literacy for GIS librarians: A discussion of languages, use cases, and competencies. *Journal of Map & Geography Libraries* **2019**, *15*, 45-67.
40. Rey, S.J. Show me the code: spatial analysis and open source. *Journal of Geographical Systems* **2009**, *11*, 191-207.

41. Rey, S.J.; Anselin, L.; Li, X.; Pahle, R.; Laura, J.; Li, W.; Koschinsky, J. Open geospatial analytics with PySAL. *ISPRS Int. J. Geo-Inf.* **2015**, *4*, 815-836.
42. Kanade, A.; Sanyal, A.; Khedker, U.P. Validation of GCC optimizers through trace generation. *Software: Practice and Experience* **2009**, *39*, 611-639.
43. Ortin, F.; Quiroga, J.; Rodriguez-Prieto, O.; Garcia, M. An empirical evaluation of Lex/Yacc and ANTLR parser generation tools. *Plos one* **2022**, *17*, e0264326.
44. He, X.; Zhang, T.; Pan, M.; Ma, Z.; Hu, C.-J. Template-based model generation. *Software & Systems Modeling* **2019**, *18*, 2051-2092.
45. Syriani, E.; Luhunu, L.; Sahraoui, H. Systematic mapping study of template-based code generation. *Computer Languages, Systems & Structures* **2018**, *52*, 43-62.
46. Balog, M.; Gaunt, A.L.; Brockschmidt, M.; Nowozin, S.; Tarlow, D. Deepcoder: Learning to write programs. *arXiv preprint arXiv:1611.01989* **2016**.
47. Zhong, V.; Xiong, C.; Socher, R. Seq2sql: Generating structured queries from natural language using reinforcement learning. *arXiv preprint arXiv:1709.00103* **2017**.
48. Alon, U.; Brody, S.; Levy, O.; Yahav, E. code2seq: Generating sequences from structured representations of code. *arXiv preprint arXiv:1808.01400* **2018**.
49. Feng, Z.; Guo, D.; Tang, D.; Duan, N.; Feng, X.; Gong, M.; Shou, L.; Qin, B.; Liu, T.; Jiang, D. Codebert: A pre-trained model for programming and natural languages. *arXiv preprint arXiv:2002.08155* **2020**.
50. Dakhel, A.M.; Majdinasab, V.; Nikanjam, A.; Khomh, F.; Desmarais, M.C.; Jiang, Z.M.J. Github copilot ai pair programmer: Asset or liability? *Journal of Systems and Software* **2023**, *203*, 111734.
51. Liang, J.; Zhao, A.; Hou, S.; Jin, F.; Wu, H. A GPT-enhanced framework on knowledge extraction and reuse for geographic analysis models in Google Earth Engine. *International Journal of Digital Earth* **2024**, *17*, 2398063.
52. Zan, D.; Chen, B.; Zhang, F.; Lu, D.; Wu, B.; Guan, B.; Wang, Y.; Lou, J.-G. Large language models meet nl2code: A survey. *arXiv preprint arXiv:2212.09420* **2022**.
53. Zhang, Y.; He, Z.; Li, J.; Lin, J.; Guan, Q.; Yu, W. MapGPT: an autonomous framework for mapping by integrating large language model and cartographic tools. *Cartography and Geographic Information Science* **2024**, *51*, 717-743.
54. Lin, Q.; Hu, R.; Li, H.; Wu, S.; Li, Y.; Fang, K.; Feng, H.; Du, Z.; Xu, L. ShapefileGPT: A Multi-Agent Large Language Model Framework for Automated Shapefile Processing. *arXiv preprint arXiv:2410.12376* **2024**.
55. Akinboyewa, T.; Li, Z.; Ning, H.; Lessani, M.N. GIS Copilot: towards an autonomous GIS agent for spatial analysis. *International Journal of Digital Earth* **2025**, *18*, 2497489.
56. Chen, L.; Guo, Q.; Jia, H.; Zeng, Z.; Wang, X.; Xu, Y.; Wu, J.; Wang, Y.; Gao, Q.; Wang, J. A survey on evaluating large language models in code generation tasks. *arXiv preprint arXiv:2408.16498* **2024**.
57. Marcilio, D.; Bonifácio, R.; Monteiro, E.; Canedo, E.; Luz, W.; Pinto, G. Are static analysis violations really fixed? a closer look at realistic usage of sonarqube. 2019; pp. 209-219.
58. Kanoutas, T.; Karanikiotis, T.; Symeonidis, A.L. Enhancing Code Readability through Automated Consistent Formatting. *Electronics* **2024**, *13*, 2073.
59. Aleissaee, A.A.; Kumar, A.; Anwer, R.M.; Khan, S.; Cholakkal, H.; Xia, G.-S.; Khan, F.S. Transformers in remote sensing: A survey. *Remote Sens.* **2023**, *15*, 1860.
60. Toth, C.; Jóźków, G. Remote sensing platforms and sensors: A survey. *ISPRS Journal of Photogrammetry and Remote Sensing* **2016**, *115*, 22-36.
61. Zhang, Q.; Gao, S. Automating Geospatial Analysis Workflows Using ChatGPT-4. 2024; pp. 715-716.